\documentclass[10pt]{iopart}

\usepackage[utf8]{inputenc}
 \expandafter\let\csname equation*\endcsname\relax
 \expandafter\let\csname endequation*\endcsname\relax

\usepackage{graphicx}
\usepackage{amsmath,amssymb}
\usepackage{dcolumn}
\usepackage{bm}

\begin{document}

\title{Charged black holes in expanding Einstein-de Sitter universes}

\author{Manuela G. Rodrigues and Vilson T. Zanchin}

\address{Centro de Ci\^encias Naturais e Humanas, Universidade Federal do
ABC, Avenida dos Estados, 5001, 09.210-580, Santo Andr\'{e}, SP, Brasil}

\begin{abstract}

Inspired by a previous work by McClure and Dyer ({\it Class. Quantum Grav.}
{\bf 23}, 1971 (2006)), we analyze some solutions of the Einstein-Maxwell
equations which were
originally written to describe charged black holes in cos\-mological
backgrounds. A detailed analysis of the electromagnetic sources for a 
sufficiently general metric is performed, and then we focus on deriving the 
electromagnetic four-current as well as the conserved electric charge of each 
metric. The charged McVittie solution is revisited and a brief study of 
its causal structure is performed, showing that it may
represent a charged black hole in an expanding universe, with the  black hole
horizon being formed at infinite late times. Charged versions of solutions
originally put forward by Vaidya (Vd) and Sultana and Dyer (SD) are also 
analyzed. It
is shown that the charged Sultana-Dyer metric requires a global electric
current, besides a central (spherically symmetric) electric charge. With 
the 
aim of comparing to the charged McVittie metric, new charged solutions of 
Vd and SD types are considered. In these cases, the original 
mass and charge parameters are replaced by particular functions of the 
cosmological time. 
In the new generalized charged Vaidya
metric the black hole horizon never forms, whereas in the new
generalized Sultana-Dyer case
both the Cauchy and the black hole horizons develop at infinite late times. 
A charged version of the Thakurta metric is also studied here. It is also a 
new solution. As in the
charged Sultana-Dyer case, the natural source of the electromagnetic field is
a central electric charge with an additional global electric
current. The global structure is briefly studied and it is verified that the
corresponding spacetime may represent a charged black hole in a cosmological
background. All the solutions present initial singularities as found in the
McVittie metric. 


\end{abstract}

\maketitle\thispagestyle{plain}

\section {Introduction}

\subsection{Preliminary remarks}

A major problem in general relativity and cosmology is related to the effects
of the cosmological expansion on gravitating local systems, such as the
solar system, accompanying the expansion of the universe. This question has
been the subject of studies since 1933, with McVittie \cite{mcvittie}.
Several models 
have emerged since then, and what is clear so far is that the way the local
dynamics is affected strongly depends upon the choice of the specific metric
to represent the system. Small scale gravitating systems, small
compared to the Hubble radius, participate in the expansion, but
the effect of the expansion being so small that can be neglected. In
such cases, the global dynamics is well described by the
Friedman-Lama\^itre-Robertson-Walker metric (FLRW), while the local dynamics
is independent of the cosmological expansion and described by an independent
metric. For large scale structures compared to the Hubble radius, the cosmic
expansion becomes significant and cannot be neglected (see,
e.g., \cite{faraoni-jacques}). In this situation, a fully dynamical metric
including the effects of the cosmic expansion must be used. In the case of a
local electromagnetic system, the analysis of \cite{price} shows
that a strongly connected system cannot be disturbed by the cosmic expansion.
Based on this, we can say that in an expanding FLRW
universe only weakly bounded systems take part in the cosmic expansion
\cite{faraoni-jacques}. However, it is still not clear what the effects
of the cosmological expansion are on other, maybe more subtle, features of 
local
gravitational systems such as the existence of trapping horizons in a black
hole spacetime.

On the other hand, the description of black hole dynamics is not well
settled; some points have
been set aside as a first approximation to simplify the analysis.
Although the black hole thermodynamics relies on the discovery of Hawking
radiation \cite{hawking,hawking2}, to avoid 
further difficulties the  backreaction has been ignored in many studies 
about black hole dynamics in cosmic backgrounds.
More complication arises when the black hole is immersed in a cosmic content
other than a cosmological constant, as dark matter and dark energy
dynamic fields (see, e.g., \cite{saida-harada-maeda} for an analysis of
Hawking radiation from black holes asymptotic to the Einstein-de Sitter
Universe). In this way, if the observations indicate $\omega<-1$, the
possibility of the Big Rip provides another motivation for the study of local
dynamics of black holes, because the event horizon can follow the expansion 
and disappear, exposing the central singularity \cite{faraoniDE}. In 
general, the cosmic content tends to be incorporated into the
black hole \cite{acrecao}, and neglecting the accretion when modeling the 
black hole evolution results in unrealistic models.

Recently observational evidence has accumulated, mainly by
the observations of SNe Ia, that sets the equation of state parameter as
$\omega=-1.02^{+0.13}_{-0.19}$ \cite{riess}, showing that the universe goes
through an accelerated expansion phase. This fact brought even more interest
on the studies on the effects of the cosmic expansion on local gravitating
systems. Therefore, it would be very useful to have exact solutions of the
Einstein equations which could describe objects with strong gravitational
field embedded in an accelerated expanding universe. Further study on the
already known exact solutions to acquire a full understating of the
geometric and physical properties of such solutions is also an
important task. This is one of the aims of the present work.

\subsection{McVittie and Vaidya type solutions}

The first known solution of Einstein equations that has been written as an
attempt to describe a spherically symmetric pointlike particle in an
expanding universe is the McVittie metric \cite{mcvittie}. In fact, 
such a metric may be used to investigate the effects of cosmic expansion on
local systems. It  is usually written in the form
\begin{equation}\begin{split} 
 ds^{2}= & \; a^{2}(t)\left(1+\frac{m}{2ra(t)}
\right)^{4}\left(d r^{2}+ r^{2}d\Omega^{2}\right)
-\left(1-\frac{m}{2r a(t)}\right)^2\left(1
+\frac{m}{2ra(t)}\right)^{-2}dt^{2},
\end{split} \label{mcvittie-metr}
\end{equation}
where $t$ is the cosmological time, $r$ is a radial comoving coordinate and
$a(t)$ is to be interpreted as the cosmological scale factor. It was first
though that metric \eqref{mcvittie-metr} could describe the Schwarzschild
black hole geometry in a cosmological
background. In this scenario the mass $m$ of the
point mass (or the central black hole) is constant, there is no flow of
energy towards the central object, or escaping from it. This non-accretion
condition
says also that the size of the central object does not change during the
expansion. However, it has been argued (see, 
e.g.,~\cite{nolan1,nolan2,nolan3}, see also
\cite{faraoni:2009uy,faraoni:2009tx}) that the McVittie metric cannot
describe a black hole evolving in an FLRW universe due to the presence of a
spacelike singularity at $ r=m/2a(t)$, which corresponds to an infinite
pressure of the cosmic fluid surrounding the object. According to these
authors, the metric describes a
singular central spherical object, but the event horizon never develops. A
timelike limiting surface and an event horizon are formed, but both shrink
due to expansion. On the other hand, it is known that in the Einstein-de
Sitter
universe, in which $a(t)\sim e^{H_0t}$, with constant $H_0$, the McVittie
metric reduces to the Schwarzschild-de Sitter metric, and the central object
described by this metric is a black hole. In fact, more recently it has been
shown that if the cosmological expansion is such that for late times the
Hubble parameter $H(t)\equiv \dot{a}(t)/a(t)$ (with the over dot 
$(\dot{\, \,})$ standing for the total derivative) reaches a constant value,
in which the spacetime
asymptotes to the $\Lambda$CDM cosmology, an event horizon is formed and
black hole and white hole regions are present in McVittie solution
\cite{kaloper,lake,faraoni:2012,fontanini}.

Straightforwardly from the McVittie metric, one can build new black hole 
type solutions in cosmological backgrounds by replacing the mass 
parameter $m$ by a function of the cosmic time $t$, $m\rightarrow m(t)$. For 
instance, by taking $m=M\,a(t)$, with constant $M$, and defining a new radial
coordinate $R$ by $R= r\left(1+\dfrac{M}{2r}\right)^2$ in 
metric~\eqref{mcvittie-metr}, we find 
\begin{equation}\label{metricTh0}
ds^{2}=    a^2(t)\left(\dfrac{dr^{2}}{1-\frac{2M}{R}}+ 
R^{2}d\Omega^{2}\right) -\left(1-\dfrac{2M}{R}\right)dt^2,
\end{equation}
which is a particular case of the Thakurta metric~\cite{thakurta}. Of 
course, other choices of $m(t)$ yield different geometries that may be 
interesting as candidates for representing black holes immersed in 
cosmological spacetimes (see below). Since the Einstein equations are not 
invariant under the change $m\rightarrow m(t)$ one sees that such a
procedure changes the sources. In the McVittie solution the source is a 
perfect fluid with energy density $\rho$ given by $8\pi\rho = 
3H^2(t)$, and pressure $p$, by $8\pi p = 
-3H^2(t) -2\dot H(t) \left[1+ m/2ra(t)\right]/\left[1-m/2ra(t)\right]$. In 
turn, for the Thakurta metric \eqref{metricTh0} the simplest 
source is a fluid with density $8\pi\rho = 
 H^2(t)/\left(1-2M/R\right)$, pressure $8\pi p = 
-\left(3H^2(t) + 2\dot 
H(t)\right)/\left(1-2M/R\right)$, and with an energy flux ${\cal Q}$ 
given by $4\pi {\cal Q} = 
M\,{H(t)}/R^2{\left(1-2M/R\right)^{3/2}}$ in the radial direction. Both 
spacetimes have singularities: McVittie at $r=m/2a(t)$ and Thakurta at 
$R=2M$.

With the aim of describing black holes in expanding universes, other metrics
have been written down. Vaidya \cite{vaidya} proposed a metric
to describe a Kerr black hole immersed in a cosmological background, which,
in the case with no rotation, can be written in the form
\begin{equation}
ds^{2} =  a^2(t)\left[dr^2 +r^{2}d\Omega^{2}\right]
+ \frac{2 m}{r}\left[\frac{d t}{a(t)}+
dr\right]^2  -d t^{2} , \label{vaidyametr-neutral}
\end{equation}
where $m$ is a constant. Vaidya argued that the metric
represents a black hole whose event horizon is located at $r a(t) = 2m
$. However, this seems a more subtle question than initially believed, since
the surface $r=2m/a(t)$ does not look like an apparent
horizon (see , e.g., \cite{mcclure:2006kg}).

A similar (Vaidya type) metric as a possible representation of a Kerr black
hole embedded
in the Einstein-de Sitter universe was examined by Thakurta \cite{thakurta}. 
As it was shown in  \cite{mcclure:2006kg}, in the Vaidya and also in
Thakurta solutions some of the energy conditions are
violated. Even though these
possible drawbacks, more studies on these metrics are necessary in order to
understand the complete geodesic structure of the corresponding spacetimes.

Another class of solutions found by Sultana and Dyer (SD) \cite{sultana} is
also of interest for the present work. The SD metric,
\begin{equation}\label{metricSD0}
ds^{2}=a^2(t)\left[dr^{2}+r^{2}d\Omega^{2}\right]
+\frac{2m}{r}\left[{dt} +{a(t)}dr\right]^{2}-dt^2,
\end{equation}
with $m$ being a constant, was proposed to represent a Schwarzschild black
hole in an Einstein-de Sitter universe. This solution is similar to the
McVittie spacetime, but with the no-accretion condition being released, i.e.,
in the SD solution there is matter (energy) accretion onto (from) the 
central object.
Another difference is that the McVittie solution can be sourced by a single
perfect fluid, while SD spacetime needs a mixture of two non-interacting
perfect fluids, a massive dust and a null (lightlike) dust fluid. In
SD solution, both in radiation and matter
dominated era, some of the energy conditions are violated.

Notice that the SD metric~\eqref{metricSD0} can be obtained from the Vaidya 
metric~\eqref{vaidyametr-neutral} by replacing the mass parameter $m$ by 
$m\,a^2(t)$. In other words, the Vaidya and the SD metric differ from each 
other only by the rate of mass accretion onto the central body.

Finally we mention here the attempts made more recently as, for instance, the
model proposed in  \cite{gao-faraoni} in which the mass of the central
object changes due to accretion of a phantom fluid, during the cosmic
expansion, considering backreaction. This metric differs from the solution
of \cite{acrecao}; however the results can not be compared
effectively since the later does not consider backreaction.

\subsection{Charged black holes in expanding universes}

 The charged black hole generalization of the McVittie metric was found 
by Vaidya and Shah \cite{vaidya1,vaidya2}. Later, a
generalization of the Vaidya~\cite{vaidya} solution including electric 
charge in the source was written~\cite{patel}.
 More recently, a very
interesting charged solution was put forward by Kastor and
Traschen~\cite{kastortraschen}. Starting with the Majumdar-Papapetrou
multi-black holes static solution of \cite{hartlehawking}, the authors
found a solution of the Einstein-Maxwell equations which describes the
dynamics of a system of charged multi-black holes in a cosmological
background. Among other interesting aspects, the solution furnishes an
analytical description of the dynamics of coalescing black holes. In the case
of a single charged black hole, the solution corresponds to the extremal
Reissner-Nordstr\"om black hole in a spatially flat Einstein-de Sitter 
universe. The global
structure of the Kastor-Traschen spacetime was explored in 
\cite{nakao}. The construction of 
dynamical multi-centered black hole type solutions was generalized to an 
arbitrary FLRW universe in a recent interesting work (see 
\cite{chimento}).

 Following a different path, and long before the works on multi-centered 
charged black holes, in~\cite{vaidya1,vaidya2} it was put forward a 
charged version of the McVittie, which is interpreted as a charged 
Reissner-Nordstr\"om (RN) metric embedded in a FLRW cosmological background. 
The influence of the cosmological evolution on the size of the RN black 
hole, as well as the motion of a test particle close to central charged body,
were analyzed in \cite{gaozhang}. Several properties of such a spacetime 
were investigated, but the global structure was not explored yet.

 In  \cite{mcclure:2006kg} a set of metrics representing charged black 
hole type solutions in expanding backgrounds was displayed. The authors were 
interested mostly in verifying the energy conditions of the resulting 
spacetimes. They analyzed  different charged metrics. The first one was a 
charged version of the McVittie metric, corresponding to a particular case 
of the solution presented in~\cite{vaidya2,gaozhang}. A charged Vaidya
metric 
was also considered. Other metrics studied there which interest us here were 
the charged Thakurta~\cite{thakurta} and the charged 
SD~\cite{sultana} metrics. A new solution, similar to the McVittie 
spacetime, was also considered in \cite{mcclure:2006kg}. The energy 
conditions were studied in all of the solutions and it was shown that some 
of them may be of physical interest.

 Some of the solutions considered in \cite{mcclure:2006kg} may
be obtained from the corresponding charged static solutions through
conformal transformations, from what some
of the properties  of the resulting spacetime could
be inferred. However, other cases such as the Vaidya type metrics are not
obtained in such a way, and the sources of the electromagnetic field, as
well as the matter content of the background, must to be analyzed with care.
Our aim in this paper is firstly investigate such electromagnetic sources in
detail, and explore the main properties of the corresponding energy-momentum
tensor. Moreover, we also briefly investigate the global structure of the
associated spacetimes.

\subsection{This paper}

The main aim of the present work is to investigate the electromagnetic 
sources of some charged black hole type metrics in expanding universes. 
Inspired by the charged versions of the McVittie \cite{mcvittie}
and Vaidya \cite{vaidya} type metrics previously considered in the 
literature 
\cite{vaidya1,vaidya2,gaozhang,mcclure:2006kg}, we write a sufficiently
general metric with 
spherical symmetry and show that, under certain circumstances, there can be 
a 
global electric current in the radial direction. Then, some particular cases 
are studied. In particular, the charged Vaidya, Sultana-Dyer, and Thakurta 
metrics reported in \cite{mcclure:2006kg} are considered. The mass 
parameters of these 
solutions, as well as the charge parameters, are taken as a function of 
the cosmological time, and new charged solutions are then built.
A brief analysis of each new metric is given.

The paper is structured as follows. In Section 
\ref{sec-gen-metric} we write a sufficiently general metric and give the 
general form 
of the Faraday-Maxwell tensor, and of the stress-energy (energy-momentum) 
tensor of the electromagnetic field related to such a metric. The 
definition of 
apparent horizon is also given in that section. Then, the charged McVittie 
metric is studied in section \ref{sect-mcvittie}, where attention is paid to 
the  electromagnetic aspects of the solution. Some aspects of the global 
structure of the corresponding spacetime are also considered. The studies 
reported in section~\ref{sect-mcvittie} are not new, but the 
results are presented there for comparison to the other cases.
Section \ref{sect-vaidya} is dedicated to study the general properties of a
charged metric
written based on the original work by Vaidya~\cite{vaidya}. Here we replace 
the mass parameter of the original metric by 
an appropriate function of the cosmic time. Again the main 
interest is on the electromagnetic quantities of the solution, and a brief 
analysis of the global structure of the resulting spacetime is given. The 
results about apparent horizons presented in this section are new results.
A charged version of a metric originally given by Sultana and Dyer 
\cite{sultana} is 
analyzed in section \ref{sect-SD}. It is shown that the 
electromagnetic stress-energy tensor corresponds to a situation where there 
is a global electric current across the spacetime. Again, 
the mass and charge parameters of the original solutions are replaced by 
particular functions of the cosmic time and some aspects of the 
global structure, including the apparent horizons, of this new charged 
SD spacetime are also shown. 
Similarly, a charged version of the nonrotating Thakurta metric 
\cite{thakurta} is investigated in Section \ref{sect-thakurta}. New mass and 
charge functions are considered, and some aspects 
of the global structure of the corresponding spacetime are also given. In 
Section \ref{sect-conclusion} we conclude.

\section{A sufficiently general metric, the electromagnetic field and
the stress-energy tensor}
\label{sec-gen-metric}

\subsection{The metric, Maxwell equations and solutions}
 A sufficiently general
metric for all cases we are interested here is of the form
\begin{equation}\begin{split}
ds^2 = & -f_0(r,t)\,dt^2 + a^2(t)\, f_1(r,t)\, dr^2 
     + 2\, a(t)\,f_2(r,t) dt\, dr +a^2(t)\,r^2\, f_3(r,t)\, d\Omega^2, 
                            \end{split} \label{gen-metric}
\end{equation}
where $t$ is a timelike coordinate, $r$ is a spherical coordinate, $a(t)$ is
the expansion factor, $d\Omega^2$ is the metric on the unit
sphere, and functions $f_0(r,t)$, $f_1(r,t)$, $f_2(r,t)$ and $f_3(r,t)$
depend on the indicated coordinates. To simplify notation we write
${g_1}^2(r,t)\equiv f_0(r,t)\,f_1(r,t)+{f_2}^2(r,t)$.

Let us stress that we are going to analyze a few 
particular cases of electrically charged spherically 
symmetric black hole type solutions in expanding universes, for which the 
functions $f_0$, $f_1$, $f_2$ and $f_3$ are given. 
For instance, in the case of the charged Vaidya type metrics considered in 
section~\ref{sect-vaidya}, one has the simple form $f_3(r,t) = r^2$. The 
same 
holds in the case of the SD and Thakurta type metrics considered 
below. For these three particular metrics, it results $g_1(r,t)=1$, and the 
radial coordinate $r$ plays the role of a comoving radial coordinate, 
analogously to the comoving radial coordinate in FLRW cosmologies. With this 
in mind, we present in this section an analysis of the 
electromagnetic fields and sources for the given metrics by using the 
 form given in equation~\eqref{gen-metric} to represent anyone of these 
cases. 

The general form of the Faraday–Maxwell tensor $F^{\mu\nu}$ and the 
electromagnetic stress-energy tensor for the metric~\eqref{gen-metric} are 
relevant to the present work. As a consequence of
the spherical symmetry, the only nonzero components of $F^{\mu\nu}$ are 
$F^{tr}=-F^{rt}\equiv E(r,t)$. The
contribution of the electromagnetic field to the energy-momentum tensor,
i.e.,
the electromagnetic stress-energy tensor $E_\mu^\nu =
F_{\mu\sigma}F^{\nu\sigma }/(4\pi) +
 \delta_\mu^\nu F_{\rho\sigma}F^{\rho\sigma }/(16\pi)$ 
then is
\begin{equation} \label{gen-emEMT0}
E_t^t = E_r^r =-E_\theta^\theta =-E_\varphi^\varphi
= -\dfrac{a^2(t)}{8\pi}\,{g_1}^2(r,t) E^2(r, t), 
\end{equation}
with all the other components of $E_\mu^\nu$ being identically zero. 
The Maxwell equations give the two equations
\begin{eqnarray}
& & \frac{\partial Q(r,t)}{\partial r}= 4\pi
a^3(t)\,r^{2}f_3(r,t)\,g_1(r,t)\, J^{t}(r,t), \label{gen-Max1}\\
& & \frac{\partial Q(r,t) }{\partial t}= -4\pi\,
a^3(t)\,r^{2}f_3(r,t)\,g_1(r,t)\, J^{r}(r,t),
\label{gen-Max2}
\end{eqnarray}
where we defined
\begin{equation}\label{chargedef1}
 Q(r,t)= a^3(t)\,r^{2}f_3(r,t)\,g_1(r,t)\,E(r,t), 
\end{equation}
 with $J^{t}(r,t)$ and $J^r(r,t)$ being the only nonzero components of the
electromagnetic current-density.
The other components of the Maxwell equations are identically zero.

Equation~\eqref{gen-Max1} gives that the quantity $Q(r,t)
=4\pi\, a^3(t) \int_0^r
r^{2}f_3(r,t)\,g_1(r,t)\, J^{t}(r,t)\, dr$  may be interpreted as total 
electric charge inside a sphere of radius $r$ at time $t$. This means that, 
using equation~\eqref{gen-Max2},
we may interpret $\partial Q(r,t)/\partial t$ as
the electric current across the spherical surface of radius $r$ at time $t$.

We are mostly interested in cosmological solutions with no net electric 
charge distribution across the whole spacetime. However, the location of the 
charged object cannot be determined without specifying the explicit 
form of the metric functions in equation \eqref{gen-metric}, and, moreover, 
since there must be 
some electric charge somewhere in the spacetime, the nonstationary character 
of the metric implies there exist some charge flux, i.e., a current-density 
which, of course, depends on the choice of the coordinates.
If one takes $J^t(r,t)=0$, one has from equation~\eqref{gen-Max1} 
$Q(r,t)=Q(t)$, 
a function of time alone. 
The radial 
component of the current-density $J^r(r,t)$ may be nonzero.
Therefore, there are at least two different solutions to
equations~\eqref{gen-Max1} and
\eqref{gen-Max2} which are interesting for the present analysis. One of
which holds in a region of the spacetime where there are no
electromagnetic sources, $J^{t}(r,t)=0$ and $J^r(r,t)=0$, while the
second is when electric current is allowed, i.e., for a special choice of 
coordinates, one may choose $J^{t}(r,t)=0$
but with $J^r(r,t)\neq0$. In the next sections, when studying charged 
versions of McVittie, Vaidya, SD and Thakurta metrics, we return 
to this  subject.

We can determine the total electric charge inside a spherical surface at a
fixed radial coordinate $r$ for the spacetimes represented by
metric~\eqref{gen-metric}. Let $S$ represent the spherical
surface $(t,\,r)=$ constants. The electric charge may be defined as
\begin{equation}\label{gen-charge}
 Q (r,t) = \int_S F_{\mu\nu} t^\mu n^\nu dS\, ,
\end{equation}
where $t^\mu$  is the unit normal to surfaces of constant $t$, $n^\nu$ is the
spacelike unit vector orthogonal
to $S$ pointing outwards, and $dS$ is the area element on $S$. Vectors
$t^\mu$ and $n^\mu$
also satisfy the relations $t^\mu t_\mu=-1$, $t^\mu n_\mu=0$, and
$n^\mu n_\mu=1$. In the case of metric \eqref{gen-metric}, we may
choose $t^\mu = \delta^\mu_t\sqrt{f_1(r,t)}/g_1(r,t) - \delta^\mu_r
f_2(r,t)/[a(t)g_1(r,t)\sqrt{f_1(r,t)}]$ and $n^\mu = \delta^\mu_r
/\sqrt{f_1(r,t)}$, and the resulting expression coincides with
equation~\eqref{chargedef1}. Such a definition of electric 
charge is applied to some particular cases below.

\subsection{The scale factor a(t)}
\label{sec-factors}

For the charged McVittie, Vaidya, SD and Thakurta type metrics 
which are considered in the present work, the function $a(t)$ in 
equation~\eqref{gen-metric} corresponds to the cosmological expansion 
factor.  In 
general, this function is
defined by the matter content of the universe, after solving the full
Einstein-Maxwell system of equations. However,
regarding the kind of study being performed here we satisfy ourselves by 
taking two particularly simple cases, defined below.

The first choice is, for comparison, the expansion factor used 
in~\cite{lake,fontanini}
\begin{equation}
 a(t)= \left[\sinh\left(\frac{3
k\,t}{2}\right)\right]^{2/3},
\label{factor1}
\end{equation}
with $k$ being a constant to be chosen appropriately. This choice has an
initial power-law expansion with $a(t)\sim t^{2/3}$ and a final de Sitter
accelerated phase. From now on this choice is called the case (a).

The second choice is a power-law expansion,
\begin{equation}
 a(t)= t^{\alpha}, \label{factor2}
\end{equation}
with constant $\alpha$, which corresponds to a universe filled with a perfect
fluid whose equation of state is of the form $p = \omega\, \rho$, with
$\omega = (2-3\alpha)/2$. Our choice shall be $\alpha = 2/3$, and so
$\omega=0$, a cold dark matter dominated model. This is called the case (b).

\subsection{Singularities and horizons}

 Besides the study on the electromagnetic sources, we investigate the 
curvature singularities and horizons of the chosen metrics.

  The interest on curvature singularities in the cosmological scenario is
because they are physical, i.e., true spacetimes singularities, meaning that
physical quantities such as the energy density or pressure of the
cosmological fluid diverge at the singularity. It is then
necessary to calculate the usual
curvature invariants which are built from the Riemann tensor 
$R_{\mu\nu\rho\sigma}$. As argued in \cite{kaloper}, for a
metric of the form given by
equation~\eqref{gen-metric}, the Ricci (${\cal R}$) and the Kretschmann 
(${\cal
K}$) scalars are necessary
to identify the spacetime singularities. These are
calculated for each metric investigated in the present work (see below).

In order to analyze the global properties  of a spacetime 
whose metric has the
form of equation~\eqref{gen-metric}, it is useful to 
perform a coordinate transformation which brings in a new radial
coordinate $R$ defined by
\begin{equation}
 R = a(t)\,r\, \sqrt{f_3(r,t)}. \label{R-coord}
\end{equation}
 With this, metric~\eqref{gen-metric} assumes the form
\begin{equation}
\begin{split}
ds^2 = & -\left[h(R,t) - \frac{C^2(R,t)}{A(R,t)}\right]\,dt^2 + A(R,t)\, 
dR^2 - 2C(R,t)dR\, dt  + R^2\, d\Omega^2, 
   \end{split}                         \label{gen-metric2}
\end{equation}
where the functions $A(R,t)$, $h(R,t)$ and $C(R,t)$ are given in terms of 
the original functions $f_0(r,t)$, $f_1(r,t)$, $f_2(r,t)$ and $f_3(r,t)$,
but we do not write such relations here.

At this point it is 
interesting to find the equations for the
expansions of the outgoing ($+$) and
ingoing (-) congruences of null geodesics, $\theta_\pm= \nabla_\mu k_{\pm}
^\mu$, where $k_{\pm}^\mu$ are tangent vectors to
the radial null geodesics. Using this definition and the geodesic equations 
for the metric \eqref{gen-metric2}, it follows
\begin{equation} \label{expansion}
{\theta_\pm} = 
\frac{2}{R}\left[\frac{C(R,t)}{{A(r,t)}}
\pm \sqrt{ \frac{h(R,t)}{A(R,t)}\,}\right]\left(\frac{dt}{dp}\right)_\pm ,
\end{equation}
where $\left(\dfrac{dt}{dp}\right)_\pm = k_{\pm}^t$, $p$ being an affine
parameter along the geodesic curve.
For well behaved $k_{\pm}^t $, one then sees that the zeros of 
${\theta_\pm}$ are  given by the zeros of the function between brackets  
in equation~\eqref{expansion}.
This means that surfaces for which the product $\theta_+\theta_-=0$, if 
they exist, are located at regions of the spacetime satisfying the 
equation 
\begin{equation}
 F_H(R,t) = \frac{h(R,t)}{A(R,t)}\left( 
\frac{C^2(R,t)}{h(R,t)\,A(R,t)}-1\right) = 0. \label{chi}
\end{equation}
According to Nolan~\cite{nolan2}, which follows~\cite{hayward}, for a 
spherically symmetric metric, the zeros of $F_H(R,t)$, which are in fact the 
zeros of the function $\chi(R,t) = \left(\nabla_\mu R \right)\nabla^\mu R$, 
defines a trapping boundary. For the kind of metrics we are interested here, 
these are also trapping horizons. On the other hand, other authors (see, 
e.g., \cite{faraoni:2009uy, kaloper, faraoniDE}) call such surfaces 
generically as apparent horizons. We follow the 
nomenclature of~\cite{faraoni:2009uy,kaloper}.

We can also verify that the zeros of the expansions $\theta_\pm$ coincide 
with stationary points (regions) of the 
radial null geodesics of the metric \eqref{gen-metric2}. In fact, the radial 
null geodesics are given by solutions of the 
equation
\begin{equation}
\left(\frac{dR}{dt}\right)_\pm 
=\frac{1}{A(R,t)}\left( C(R,t)\pm\sqrt{h(R,t)\, A(R,t)}\right)  ,
\end{equation}
where the plus (minus) sign indicates outgoing (ingoing) geodesics.
We see also that the functions $\left(dR/dt\right)_\pm$ change sign at the
zeros of the function $F_H(R,t)$, meaning that the zeros may be
apparent horizons separating the spacetime into trapped and non-trapped 
(regular) regions.

It is worth mentioning that, for the metric of equation~\eqref{gen-metric2}, 
one 
has $ F_H(R,t) = h(R,t)g^{RR}$, with $g^{RR}$ being the component of the 
inverse metric. Indeed, this procedure of looking for the zeros of the 
function $g^ {RR}$ to find the possible apparent horizons of black hole type 
metrics in expanding spacetimes was followed by many authors (see, e.g., 
\cite{nolan2, kaloper,lake,fontanini} for McVittie spacetime, \cite{nakao} 
for Kastor-Traschen solution \cite{kastortraschen}, and 
\cite{faraoni:2009uy} 
for a Thakurta type metric).

\subsection{About the next sections}

In the following we investigate a few particular charged metrics written to
describe charged black holes in expanding universes.
The main aim is to understand the physical nature of the electromagnetic
sources of the corresponding spacetimes.
Another interest is to look for apparent horizons in each case, the ultimate
goal is to determine if black hole
 horizons are really present, besides finding the spacetime
singularities in order to identify the physical nature of the central object
in each case. 
In order to do that, the complete geodesic structure of the
given spacetime must be investigated. However, this study is lengthy and,
moreover, all the metrics we consider here deserve a separate detailed
analysis because of the interest on the subject in recent literature. Hence,
to avoid a very long paper, in this work we investigate only the form of
the apparent horizons and the curvature singularities for each one of the
studied metrics.

\section{Charged McVittie metric}
\label{sect-mcvittie}

The generalized version of McVittie metric including the electric charge
of the central body was put forward in  \cite{vaidya2} (see also
\cite{gaozhang}). Here we review
the electromagnetic sources of such a metric, and present a short analysis
of curvature singularities and apparent horizons.

\subsection{The solution}

In the case of
an asymptotically spatially flat FLRW spacetime, the charged McVittie metric
may be obtained from the general form \eqref{gen-metric} by setting
$f_0(r,t)=
f^2(r,t)/g^2(r,t)$, $f_1(r,t) = f_3(r,t)=g^2(r,t)$, and $f_2(r,t) =0$, which
gives $g_1(r,t)=f(r,t)$, where the functions $f(r,t)$ and $g(r,t)$ are
defined respectively by
\begin{eqnarray}
 & &f(r,t) = 1-\dfrac{m^2}{4a^2(t)\, r^2}+ \dfrac{q^2}{4a^2(t)\,r^2},
 \label{f(r,t)}\\
& & g(r,t) = \left(1+\dfrac{m}{2a(t)\,r}\right)^2 - \dfrac{q^2}{4a^2(t)\,
r^2}, 
\label{g(r,t)}
\end{eqnarray}
with $m$ and $q$ being constants related respectively to the mass and  
charge of the central body. The resulting metric is
\begin{equation} \label{chargedMcVmetr}
 ds^{2}= -\frac{f^2(r,t)} {g^2(r,t)} dt^{2} +a^{2}(t)
g^2(r,t)\left(d r^{2}+ r^{2}d\Omega^{2}\right).
\end{equation}
The function $a(t)$ is interpreted as the expansion
factor with $t$ being a cosmological time. The asymptotic metric
($r\rightarrow \infty $)
results $ds^2= -dt^2 + a(t)^2\left(dx^2+ x^2d\Omega^2 \right)$, which is the
FLRW metric in the case of flat three space. The above metric suffers
from the same illness as the uncharged McVittie metric,
 but at least when the Hubble parameter $H(t)\equiv \dot{a}(t)/a$ asymptotes
a constant
at $t\rightarrow\infty$, such a solution has an event horizon and represents
a black hole in an expanding universe (see, e.g., \cite{lake, fontanini}).

Notice that a further generalization of the charged McVittie metric 
\eqref{chargedMcVmetr} is obtained simply by replacing the constant 
parameters $m$ and $q$ by function of the cosmological time, $m(t)$ and 
$q(t)$. As discussed in the literature (see e.g. 
\cite{nolan1,nolan2,faraoni:2009uy}), this implies in the violation of the 
non-accretion hypothesis of McVittie, and there must be some kind of energy 
flux in the radial direction throughout the spacetime. In respect to the 
electric charge, it means a non-constant electric charge of the source, and 
the presence of a radial electric current throughout the spacetime (see 
also the next 
section).

\subsection{Faraday and stress-energy tensors, conserved electric charge and
current-density}
\label{sec-Faraday-McV}

The electromagnetic source and the stress-energy tensor for the
metric~\eqref{chargedMcVmetr} were analyzed in~\cite{vaidya2,gaozhang}
and then
we do not need to reproduce the results here. However, further comments
with respect to the Maxwell equations are in order. According to our analysis
in section~\ref{sec-gen-metric}, the nonzero components of the
Faraday-Maxwell strength tensor are $F^{tr}=-F^{rt}\equiv E(r,t)$, and
for the charged McVittie metric one finds (see also
\cite{vaidya2,gaozhang})
$ 
 E(r,t) = \dfrac{q_0 h(t)}{a^3(t)\,r^2f(r,t)\, g^2(r,t)}.
$
The solution presented in \cite{vaidya2} assumes $h(t)=1$, leading to
$Q(r,t)=q_0= {\rm constant}$, and $q_0$ is a conserved
quantity identified with the total electric charge of the
source, so that $q_0=q$, $q$ being the charge parameter of the metric. This
result means that
comoving observers in the charged McVittie spacetime \eqref{chargedMcVmetr}
attribute a constant electric charge to the central object.
Furthermore, with such a choice, the Einstein-Maxwell equations in the
presence of a cosmic perfect fluid are satisfied resulting in
Friedmann type equations for the energy-density and pressure. No additional
terms in the density or pressure arise because of
the presence of the electric charge.

 On the other hand, if one chooses $h(t)\neq 1$, the Maxwell equations are
satisfied with the inclusion of a radial current-density given by
 $J^r(r,t)=-\dfrac{q_0\,\dot h(t)}
{ 4\pi\,r^2a^3(t)\, g^2(r,t)}$, and the charge of the central object is not
constant. In addition, in such a case there would be additional
contributions to the energy density and pressure of the cosmic fluid coming
from the electromagnetic field. For instance, the energy density has a term
of the form $q^2(t)/r^4 - q_0^2 h^2(t)/r^4$, which, for $h(t)\neq 1$ and 
constant $q$
implies in violation of the energy conditions. This solution has not been
investigated in the literature, since it is regarded as unphysical. It is 
worth mentioning that if we replace the parameter $q$ by a 
function of time and take $q(t)=q_0 h(t)$, the Maxwell equations are
satisfied and 
the spurious term in the energy (and pressure) of the perfect fluid vanishes.
However, in order to satisfy the Einstein equations additional matter 
components must be added to the background. A detailed study of such a case 
is not the purpose of this work.

\subsection{Singularities and horizons}

 As already mentioned, we do not investigate the complete geodesic
structure of the charged McVittie metric here, since a complete analysis is
lengthy and out of the scope of this paper. However, the results on the
singularities and horizons are of interest. We consider here just the case 
of constant mass and charge parameters $m$ and $q$, and show the 
singularities and apparent horizons of metric 
\eqref{chargedMcVmetr} for comparison to the other solutions investigated in 
the next sections. 
More details about the apparent horizons and singularities in the case of 
constant parameters $m$ and $q$ can be found in \cite{sodaiv}.

To say something about curvature singularities of the charged McVittie
metric~\eqref{chargedMcVmetr}, we calculate the
Ricci and the Kretschmann scalars (see \ref{appendixMcV}).
To simplify expressions it is convenient to use a new radial
coordinate $R$ defined by $R = a(t)\,r\,g(r,t)$ (see \cite{gaozhang}).
The Ricci scalar is then ${\cal R}=12 H^2(t)+ 6\dot H(t)/\sqrt{1-2M/R +
Q^2/R^2}$, where we have written $M=m$ and $Q = q$. The Kretschmann
scalar is written in \ref{appendixMcV}. We immediately see possible
singularities when $H(t)$ diverges, and at the region of the spacetime
 where $h(R)=1-2M/R + Q^2/R^2=0$, i.e., for $R= M \pm\sqrt{M^2-Q^2}$, or, in
terms of the original coordinate $r$, at  $r\, a(t)=\pm\sqrt{M^2 -Q^2}/2$.
In the overcharged case ($M^2< Q^2$), the function $h(R)$ has no real roots
and the singularity is at $R=0$ (at $ r\,a(t)= -(M\pm Q)/2 $). We consider
here just the undercharged case ($M^2>Q^2$).

The apparent horizons of the McVittie metric are given by the roots of
the equation (see equation~\eqref{chi})
\begin{equation}
 F_M(R,t)=-H^2(t)\, R^2 +\left(1-\dfrac{2M}{R}+ \dfrac{Q^2}{R^2}\right) =0, 
\label{F-M}
\end{equation}
which is a fourth-order
polynomial equation for $R$, given the solutions in terms of the time $t$.
 Note also that the zeros of $F_M(R,t)$ coincide with the zeros of the radial
coordinate velocity $dR/dt$ of ingoing lightlike geodesics, i.e., the zeros
of $\frac{dR}{dt} = H(t)\, R\sqrt{1- \frac{2M}{R} +
\frac{Q^2}{R^2}} 
-\left(1- \frac{2M}{R} +
\frac{Q^2}{R^2}\right) $, for which $1- {2M}/{R} +{Q^2}/{R^2} \neq 0.$

\begin{figure}[!ht]
\begin{center}
\includegraphics[width=3.5in]{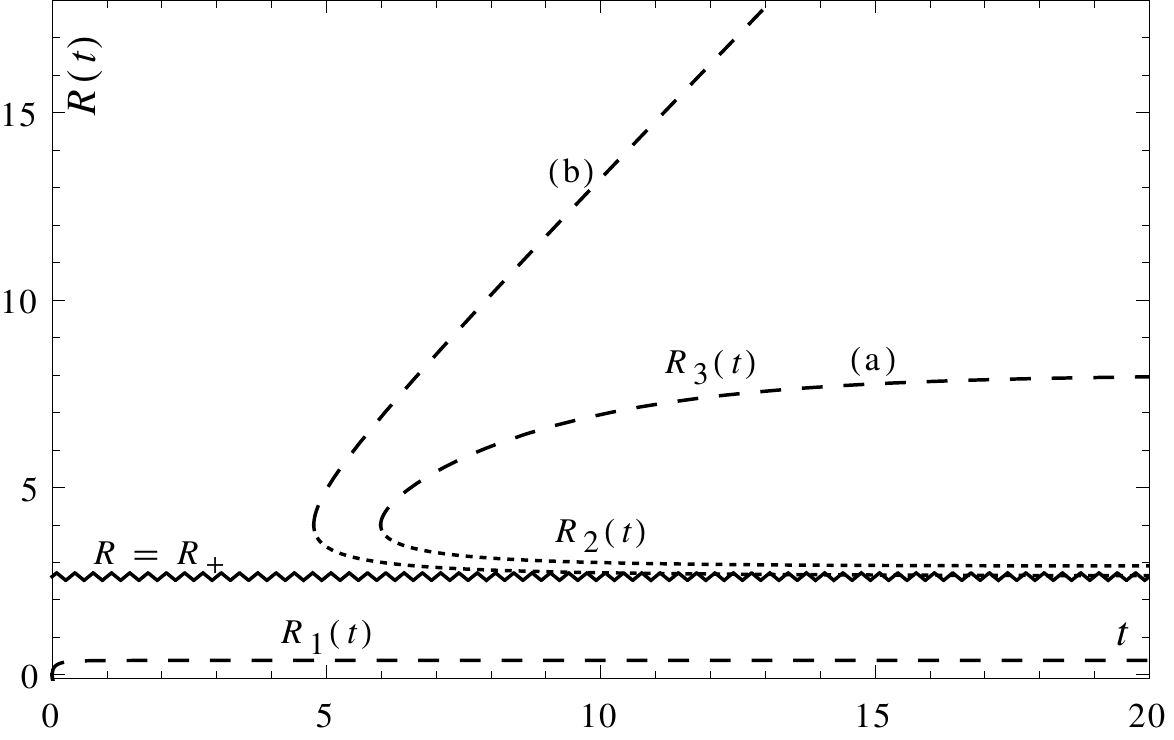}
\caption{The evolution of the apparent horizons as a function of the
time $t$ in the undercharged case of the charged McVittie metric.
We used $M=1.5,\, Q=1.0$. The case (a) is for the scale factor
of equation~\eqref{factor1} with $k=0.1$, and the case (b) for $a(t)$ given 
by
equation~\eqref{factor2} with $\alpha = 2/3$.
For sufficiently large times there
are three real positive roots of the equation $F_M(R,t)=0$, with
$F_M(R,t)$ given by equation \eqref{F-M}. There is an
initial singularity at $R=R_+= M +\sqrt{M^2- Q^2},\, t=$ finite, 
$R_1(t)$ is always smaller than $R_+$, $R_3(t)$ is a cosmological type 
horizon while $R_2(t)$ is a black hole type apparent horizon.}
\label{horizons-McVittie}
\end{center}
\end{figure}

 For the undercharged McVittie metric ($M^2> Q^2$),  and for the chosen 
form and parameters of the scale factor, for which $k\times M$ is 
sufficiently small (case (a)), and $\alpha =2/3$ (case (b)), function 
$F_M(R,t)$ has
just one real root at small times, while it has three real roots, 
$R_1(t)$, $R_2(t)$ and $R_3(t)$ for sufficiently large times (see
figure~\ref{horizons-McVittie}). The 
root $R_1(t)$ is always inside the
singularity radius $R_+= M + \sqrt{M^2-Q^2}$ and so it is of no interest, as
long as we are interested in the $R> R_+$ region. In view of the similarity
between functions $R_2(t)$ and $R_3(t)$ shown in 
figure~\ref{horizons-McVittie}
and the corresponding functions of the uncharged McVittie case, we may infer
that the geodesic structure of the undercharged McVittie is similar to the
original McVittie spacetime, at least for $R\geq R_+$ (see e.g.
\cite{nolan1,nolan2,nolan3,kaloper,lake,fontanini}).
If the cosmological expansion is such that $\displaystyle
\lim_{t\rightarrow\infty} H(t)=H_0=$constant (case (a)), there are two
horizons in the future of the initial singularity. They correspond to the 
spherical
surfaces
$R_c$ and $R_h$ which are the asymptotic values of the functions $R_3(t)$
and $R_2(t)$ at $t\rightarrow\infty$, respectively. According to the study 
of 
\cite{lake}, the black hole horizon ${\mathcal H}$ is defined by 
($R_2,\, t=\infty$), while $R_c$ defined by ($R_3,\, t\rightarrow \infty$) 
gives a de Sitter horizon (case (a)), or the Hubble horizon (case (b)). Our 
$R_h$ here corresponds to the 
asymptotic limit of $R_ -$ of \cite{lake}. The apparent horizon $R_1(t)$ is 
inside the singularity, but it has an interesting interpretation, since it
asymptotes a charged black hole inner (Cauchy) 
horizon at $t\rightarrow\infty$, i.e., the region ($R_1,\, t=\infty$) is 
similar to the Reissner-Nordstr\"om-de Sitter inner horizon; it is a Cauchy
horizon.
 In case (b), in which $\displaystyle
\lim_{t\rightarrow\infty} H(t)=0$,   the conformal structure of the
boundary of the charged McVittie spacetime is similar to that of a
Reissner-Nordstr\"om black hole, apart from
the naked initial singularities. In this case, the asymptotic form of the
metric in $(R,t)$ coordinates is time independent. In both cases (a) and (b),
the asymptotic metric, at $t\rightarrow\infty$, can be put into stationary
forms, so that both admit Killing vectors associated to time translations. 
It is
interesting to explore the geodesic structure of the charged McVittie
spacetime in detail, including the extremal and overcharged cases. However 
we are not going to report the results of such a study here.

\section{Charged Vaidya type metrics }
\label{sect-vaidya}

A charged version of the Vaidya metric \cite{vaidya} was reported in
\cite{mcclure:2006kg}. Here we review the electromagnetic sources of a 
modified version of that metric, and present a brief analysis
of curvature singularities and apparent horizons. The modifications of the
metric is done by considering also time-varying mass and charge parameters 
when compared to the solution given 
in~\cite{mcclure:2006kg}.

\subsection{The solution}
According to our notation of
section~\ref{sec-gen-metric}, the charged Vaidya type solution given in
\cite{mcclure:2006kg} is obtained by choosing
$ f_0(r,t)= 1- g(r)/a^2(t) $,
$f_1(r,t)= 1 +g(r)/a^2(t)$, $f_2(r,t)=g(r)/a^2(t)$, and $f_3(r,t)=1$, with
$g(r)= 2m/r -q^2/r^2$, which gives $g_1(r,t)= 1$ (see 
equation~\eqref{gen-metric}).
Here we consider a generalized version of the metric given in 
\cite{mcclure:2006kg} by taking the mass and charge parameters as a 
function of time, i.e., $m=m(t)$ and $q=q(t)$, and so we write 
\begin{eqnarray}
ds^{2} = a^2(t)\!\left[dr^2 +r^{2}d\Omega^{2}\right]\!  
+ g(r,t)\!\left[\frac{d t}{a(t)}+ dr\right]^2\!-dt^{2} ,
\label{chargedvaidyametr}
\end{eqnarray}
where
\begin{equation}
 g(r,t) = \frac{2m(t)}{r} -\frac{q^2(t)}{r^2}. \label{g0(r,t)}
\end{equation}

As promptly seen from~\eqref{chargedvaidyametr}, for $a(t)=1$ and constant 
$m$ and $q$ it results
the Reissner-Nordstr\"om black hole spacetime, and far from the central body,
$r\rightarrow\infty$, it results the Einstein-de Sitter cosmological
spacetime.

The nonzero components of the Einstein tensor corresponding to
metric~\eqref{chargedvaidyametr} are the diagonal ones
${G^t}_t,\,{G^r}_r,\, {G^\theta}_\theta = {G^\varphi}_\varphi,$ besides the
ones that mix the radial and time coordinates, namely ${G^t}_r$ and
${G^r}_t$. We do not write the full expressions of these components here, 
but we observe that there are changes in the sources in comparison to the 
results given in \cite{mcclure:2006kg} because of 
the time dependence of  $m(t)$ and $q(t)$. To be 
explicit, we take, for instance, the new terms that arise in the component 
${G^t}_t$,  which are $- 2 H((t)\left(\dot m(t)a(t)r - q(t)\dot 
q(t)\right)/a^2(t)r^2$. 
As in the original Vaidya metric \cite{vaidya}, these terms 
are singular at $r=0$. Their contributions to the energy density depend on 
the explicit form of the functions $m(t)$ and $q(t)$, and in the case 
$m(t)= M/a(t)$, $q(t) = Q/a(t)$, considered in the next subsection, the 
problem of negative energy densities (at early times $t$ and large $r$) of 
the original Vaidya solution persists (see, e.g., 
\cite{mcclure:2006kg}, 
see also \ref{appendixVy}).
We need to  point out explicitly the linear terms on $q^2(t)$ and that do 
not contain derivatives of $a(t)$, because 
they are directly related to the electromagnetic stress-energy tensor, what 
is  one of our main points of interest. The
contribution of these terms to the Einstein tensor are the following
\begin{eqnarray}\label{einsTVaidya}
{G^t}_t =  {G^r}_r = -{G^\theta}_\theta = -{G^\varphi}_\varphi = 
-\dfrac{q^2(t)}{a^4(t)r^4}.
\end{eqnarray}
From the analysis of section~\ref{sec-gen-metric}, it is expected that 
terms of the form $q^2(t)/a^4(t)\,r^4$ would 
be exactly canceled by the corresponding components of the electromagnetic
stress-energy tensor $E_\mu^\nu$. This fact is confirmed in the next
subsection.

\subsection{Faraday and stress-energy tensors, conserved electric charge and
current-density}

Following the analysis of section~\ref{sec-gen-metric}, one gets the
Faraday-Maxwell tensor field as $ E(r,t)=q_0h(t)/a^3(t)r^2$, $q_0$ being a 
constant,  which yields
 $ Q(r,t)= q_0h(t)$ (see equation~\eqref{chargedef1}), and the 
electromagnetic stress-energy tensor then is
\begin{equation} \label{emEMTVaidya}
E_t^t = E_r^r =-E_\theta^\theta =-E_\varphi^\varphi  = 
\dfrac{-q_0^2h^2(t)}{ 8\pi a^4(t)\,r^4}\,,
\end{equation}
with the other components being identically zero.

It is seen that by choosing $h(t)=1$, i.e., $E(r,t)=q_0/a^3(t)r^2$, with
constant $q_0$, the Maxwell equations are satisfied with zero
current-density, $J^t(r,t)=0$, which implies that $q(t)= q_0$ is the total
conserved electric charge of the source, as expected. In fact, from Maxwell 
equations~\eqref{gen-Max1} and \eqref{gen-Max2},
we see that the electric four-current is not well defined just in the region 
of the spacetime for which $a^3(t)\, r^2 = 0$. Everywhere
else in the spacetime the electromagnetic sources vanish, and then it is 
natural to associate the loci where $a^3(t)\, r^2 = 0$ as the 
actual location of the source.
Furthermore, comparing
equations~\eqref{emEMTVaidya} with equations~\eqref{einsTVaidya}, it is seen 
that
the components of the electromagnetic stress-energy tensor
cancel the corresponding terms of the Einstein tensor through Einstein
equations ($G_{\mu\nu} = 8\pi\, E_{\mu\nu}$).

Contrarily, if one takes $h(t)\neq 1$, so that $E(r,t)=q_0h(t)/a^3(t)r^2$,
the radial electric current-density is nonzero, $J^{r} (r,t) =
-\dfrac{q_0\,\dot
h(t)}
{ 4\pi\,r^2a^3(t)}$, and the total charge of the source is not a constant.
In this case, there may arise contributions to the energy-momentum tensor
from the electromagnetic field that violate the energy conditions. In fact,
similarly to the uncharged case (as investigated, e.g., in
\cite{vaidya}), the charged Vaidya metric
\eqref{chargedvaidyametr} can be sourced by a global perfect cosmic fluid in
addition to a heat flux, besides the spherical charged object.
If $h(t)\neq 1$, the energy density (and
also the pressure) of the cosmic fluid acquires a term proportional to
$q^2(t)- q_0^2h^2(t)$.  The time component of
the Einstein equations gives $G_t^t - 8\pi E_t^t= -\left(q^2(t) -
q_0^2h^2(t)\right)/r^4a^4(t)$. This is similar to what happens in the charged
McVittie metric we investigated in subsection~\ref{sec-Faraday-McV}, if 
$ q^2(t)\neq  q_0^2h^2(t)$ the
energy conditions may be violated and the corresponding solutions is
disregarded as unphysical.

\subsection{Singularities and horizons}

Let us comment here on the curvature singularities and possible
apparent horizons of the spacetime defined by the charged Vaidya type
metric~\eqref{chargedvaidyametr}.

In the case we choose constant $m$ and $q$, the Ricci scalar is given by
$
{\cal R} =2H^{2}(t)\left(6+{g(r)}/{a^2(t)}\right)+2\dot{H}(t)\left(3+
{2g(r)}/{a^2(t)}\right)-{ 4mH(t)}/{a^{3}(t)r^{2}},
$
while the Kretschmann scalar is given in \ref{appendixVy}.
 We can note that both scalars are singular when $H(t)\rightarrow\infty$, and
at $a(t)\, r\rightarrow 0$.

\begin{figure}[!ht]
\begin{center}
\includegraphics[width=3.5in]{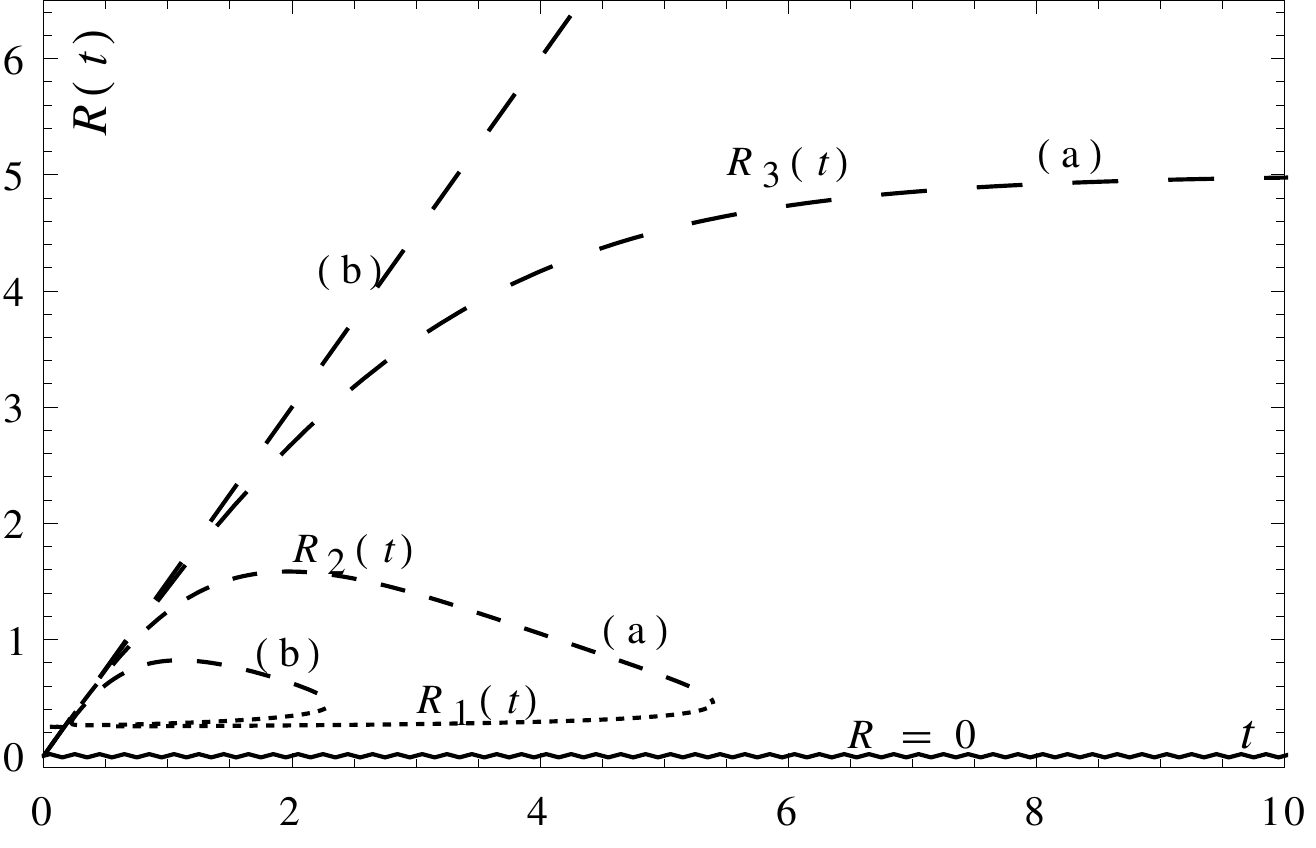}
\caption{The evolution of the apparent horizons as a function of the
time $t$ in the undercharged case of the charged Vaidya metric.
We used $M=2.0,\, Q=1.0$. The case (a) is for the scale factor 
of equation~\eqref{factor1} with $k=0.2$, and the case (b) for $a(t)$ given 
by
equation~\eqref{factor2} with $\alpha = 2/3$.
For sufficiently large times there
is only one real positive root of the equation $F_V(R,t)=0$, with
$F_V(R,t)$ given by equation \eqref{F-V},
which corresponds to a cosmological (Hubble like) horizon. There is an
initial singularity when $a(t)=0$ and also at $R=0,\, t=$ finite.}
\label{horizons-Vaidya}
\end{center}
\end{figure}
 
For the non-constant $m(t)$ and $q(t)$ case there are of course many 
interesting choices. To keep the metric similar to the McVittie one, and to 
compare with the SD and Thakurta type metrics in the following, we 
take the case $m(t) = M/a(t)$ and $q(t)= Q/a(t)$ with constant $M$ and $Q$.
A long but straightforward calculation shows that the curvature scalars are 
singular at the same regions as in the case with constant $m$ and $q$, where 
$a(t) r =0$. As commented in \ref{appendixVy}, the curvature 
scalars show that the singularities are the same as for constant $m$ and $q$.

Similarly to the case of the charged McVittie metric, the 
trapping horizons of a spacetime whose metric has the
form of equation~\eqref{chargedvaidyametr}
may be found by
analyzing the roots of the equation (see equation~\eqref{chi}),
\begin{equation}\label{F-V}
 F_{V}(R,t)\equiv H^2(t)\, R^2+ 
 \left[H(t)\, R -1\right]^2 \frac{g(R)}{a^2(t)} -1 =0, 
\end{equation}
where now $R = ra(t)$,  $g(R)={2M}/{R}-{Q^{2}}/{R^{2}}$, 
and we defined  $M=m\, a(t)$ and $Q=q\,a(t)$. Here we consider the case of 
constant $M$ and $Q$ only.

The zeros of the function $F_{V}(R,t)$ tell us where the expansion rate of
radial null geodesics changes signs, which is a necessary condition to find
apparent horizons.  The
short analysis we give here is by comparison to the McVittie metric studied
in~\cite{lake,fontanini}.
The solutions of equation~\eqref{F-V} depend explicitly on $a(t)$ and hence 
we
need to define such a function a priori. In the present analysis we use the
two cases mentioned in section~\ref{sec-factors}.

Let us first mention that equation~\eqref{F-V} has at least one real positive
root for all times given by $ H(t) R =1$. 
This solution is shown by the curves $R_3(t)$ in
figure~\ref{horizons-Vaidya}; it is the Hubble radius, it tends
asymptotically to a cosmological de Sitter horizon in
case (a), with $R= 1/H(\infty)={\rm constant}$, and is exactly $3t/2$ in 
case (b).
Moreover, since in an expanding cosmological scenario the scale
factor becomes arbitrarily large for infinite time, the second term in
equation~\eqref{F-V} vanishes and $R_3(t)$ is the only root at late times.
For simplicity we consider here the undercharged case only, $M^2 > Q^2$.
The roots $R_1(t)$, $R_2(t)$, and $R_3(t)$,
as function of the time $t$, are plotted in figure~\ref{horizons-Vaidya}
for the case with $M=2.0$ and $Q=1.0$, and for the scale factor $a(t)$ 
labeled as cases (a) and (b); see 
section~\ref{sec-factors}.
Since the asymptotic form (for infinite time) of the roots of
the function $F_{V}(R,t)$ dictates the structure of the boundaries of
the given spacetime, we see that a black hole horizon analogous to 
${\mathcal H} = (R_2,\, t=\infty)$ of the charged McVittie metric 
never forms in
the charged Vaidya metric. The black hole type apparent horizon $R_2(t)$ is 
formed after the initial singularity (at the same time as the inner 
apparent horizon $R_1(t)$), expands faster than $R_1(t)$ at early times, but 
starts to shrink after a given time while $R_1(t)$ keeps growing slowly with 
time. Finally, after a finite time interval, both apparent horizons 
coalesce and disappear.  The singularity at $R=0$ is naked for all later 
times.
Assuming, for instance, a matter dominated universe at late times, then
$\lim_{t
\rightarrow\infty} a(t)\,H(t) = 0$ (case (b)) the only real positive root
increases indefinitely with time, corresponding to the Hubble radius.
When the
asymptotic form of the scale factor $a(t)$ corresponds to an accelerated
expanding universe (case (a)), where $\lim_{t \rightarrow\infty} H(t) =
H_0$= constant, the only real positive root tends to a constant, which
corresponds to a de Sitter cosmological horizon. Even though these partial
results have been found, a more detailed study is necessary to
determine the global structure of the spacetime generated by 
metric \eqref{chargedvaidyametr}.

\section{Charged Sultana-Dyer type metrics}
\label{sect-SD}

\subsection{The solution}
A charged version of the SD metric, which is obtained through 
a conformal transformation of the Reissner-Nordstr\"om metric written in
Eddington-Finkelstein coordinates, was given in \cite{mcclure:2006kg}.
For the present analysis we take the following generalized SD type metric
\begin{equation}
ds^{2} =  a^2(t)\left[dr^2 +r^{2}d\Omega^{2}\right] 
 +  g(r,t)\left[\,d t+
a(t)\,dr\right]^2 -d t^{2} , \label{chargedSDmetr}
\end{equation}
where again we have put $g(r,t) = 2m(t)/r -q^2(t)/r^2$ (see 
equation~\eqref{g0(r,t)}).
As noted earlier, the factor $a(t)$ is interpreted as the cosmological scale 
factor,
with $ t$ being the cosmological time. This charged SD
metric is obtained from our general metric \eqref{gen-metric} through the
identifications $f_0(r,t) = 1-2m(t)/r+q^2(t)/r^2$, $f_1(r,t)= 1+ 
2m(t)/r-q^2(t)/r^2$,
$f_2(r,t)= g(r,t)= 2m(t)/r - q^2(t)/r^2$, and $f_3(r,t)=1$; from what
follows $g_1(r,t)=1$. One can see that the SD metric \eqref{chargedSDmetr}
follows from the Vaidya metric by substituting $m(t)$ and $q(t)$
respectively by $m(t)\,a^2(t)$ and $q(t)\, a(t)$ into the
metric~\eqref{chargedvaidyametr}.

The uncharged version of metric \eqref{chargedSDmetr} has been
considered in the
literature as a good candidate to represent a Schwarzschild-like black hole
immersed in an expanding cosmological background
(see e.g. \cite{sultana,faraoni:2009uy}). On the
other hand, this charged version \eqref{chargedSDmetr} presents an additional
issue that deserves to be considered carefully. For future reference let us
note that metric \eqref{chargedSDmetr} yields an Einstein tensor whose
relevant components presents terms like
\begin{equation} \label{EinstEM-SD}
{G^t}_t=
{G^r}_r= -{G^\theta}_\theta = -
{G^\varphi}_\varphi=-\dfrac{q^2(t)}{a^2(t)\,r^4}.
\end{equation}
Compared to the Vaidya metric case, equation~\eqref{einsTVaidya}, the 
difference is just by a factor of $a^2(t)$.
In the following subsection we give the Faraday-Maxwell and the
electromagnetic stress-energy
tensors for the charged SD metric and compare to
Einstein tensor~\eqref{EinstEM-SD} to analyze the source of the
electromagnetic field.

\subsection{Faraday and stress-energy tensors, conserved electric charge and
current-density}

According to the study of section~\ref{sec-gen-metric}, the Faraday-Maxwell
tensor field for the SD metric~\eqref{metricSD0} is of the form $
F^{tr}=E(r,t)=q_0h(t)/a^3(t)r^2$, where $q_0$ is a constant and $h(t)$ is an
arbitrary function of time,  and then the electric charge obtained from
equation~\eqref{chargedef1} is $Q(r,t)= q_0h(t)$. The nonzero
components of the electromagnetic stress-energy tensor is
\begin{equation} \label{emEMTSD}
E_t^t = E_r^r =-E_\theta^\theta =-E_\varphi^\varphi  = 
\dfrac{-q_0^2h^2(t)}{ 8\pi a^4(t)\,r^4}\,.
\end{equation}  
Comparing this with the Einstein tensor components 
(equations~\eqref{EinstEM-SD}) we get terms of the form 
$ G_t^t - 8\pi E_t^t =  \left({ {q_0}^2h^2(t) }-{q^2(t)\,a^2(t)}
\right)/\left(a^4(t)r^4\right)$. The exact 
cancellation  of such terms is expected, and then we must have 
$q^2(t)= q_0^2h^2(t)/a^2(t)$. 

If one takes $q(t)=q_0=$ constant, as the solution given in
\cite{mcclure:2006kg}, the expected cancellation happens only if 
$h(t) = a(t)$. However, in that case the constant $q_0=q$ is 
not the conserved electric charge. In fact, using the charge definition 
\eqref{gen-charge} we get $Q(r,t) = q_0\, a(t)$. If so, there would be a 
nonzero electric current in the radial direction $J^{r}(r,t)= 
-q_0H(t)/(4\pi a^2(t)r^2)$, implying that the total charge of the central
object varies with 
time. This is the solution considered in \cite{mcclure:2006kg}, but the 
presence of the radial electric current was not noticed there. 

On the other hand, the natural choice is to take $Q(t) = q_0=\,$ constant, 
where the total 
charge of the central body is constant (cf. equation~\eqref{gen-charge}), 
with 
$h(t)=1$, which corresponds to 
the case  of zero four-current density throughout the spacetime, $J^\mu =0$.
In such a case, one must have $q^2(t) = q_0^2/a^2(t)$, giving rise to a new 
metric of SD type. To see how this choice affects the sources let 
us take the full expression for $ G_t^t-8\pi E_t^t $ in the case with
$m(t)=M/a(t)$ and $q(t)=Q/a(t)$, where $M$ and $Q$ are constant parameters. 
This gives  $ G_t^t-8\pi E_t^t = -H^2(t)\left(3+\dfrac{4M}{a(t)r }
-\dfrac{Q^2}{a^2(t)r^2}\right)+\dfrac{4MH(t)}{a^2(t)r^2}$. In the 
original charged SD metric of \cite{mcclure:2006kg} one has $ 
G_t^t-8\pi 
E_t^t = -3H^2(t)\left(1+\dfrac{2m}{r }
-\dfrac{q^2}{r^2}\right)+\dfrac{4mH(t)}{a(t)r^2}$.  Hence, apart from the 
time dependence of the mass and charge functions, this component of the 
Einstein equations has no significant differences in comparison with the 
solution 
given in \cite{mcclure:2006kg}. Similar changes occur in the other 
components of the Einstein tensor, but we do not analyze them here.

We then conclude that the simplest case with $q(t)=q_0=\,$ constant and 
equal to the electric charge of the spherical source is physically not 
interesting. The
Einstein-Maxwell equations are
satisfied only with the presence of an additional energy-momentum terms that
may violate the energy conditions. The energy density acquires an extra
term of the form $ q_0^2\left(1- a^2(t) \right)/\left(a^4(t)r^4\right)$. 
This term changes sign at time $t$ such that
$a^2(t)= 1$, which is an arbitrary time, since the scale factor 
$a(t)$ can be normalized in such a way that 
$a^2(t)=1$ at any given time. 
A similar 
situation happens in the case of the charged Thakurta spacetime, which we 
investigate in more detail below.

\subsection{Singularities and horizons}

Here we comment briefly on the causal structure of the charged SD
spacetime given by the metric~\eqref{chargedSDmetr}. 
The Ricci and the Kretschmann scalars for this metric are written in 
\ref{appendixSD}, first for constant $m$ and $q$ and afterwards for 
$m(t)=M/a(t)$ and $q(t) = Q/a(t)$, with 
constant $M$ and $Q$. 
 As in the case of charged Vaidya solution, both scalars are singular when
$H(t)\rightarrow\infty$ and at $a(t)\, r=0$. The singularities are the same 
when the mass $m$ and the charge $q$ depend on time, as it can be seen by 
calculating the Ricci and the Kretschmann scalars for such a case.
Again, we consider here the case where $m(t)=M/a(t)$ and $q(t) = Q/a(t)$. As 
it is seen below, this choice implies the metric 
\eqref{metricSD0} has a geodesic structure similar to the charged McVittie 
metric.

\begin{figure}[!ht]
\begin{center}
\includegraphics[width=3.5in]{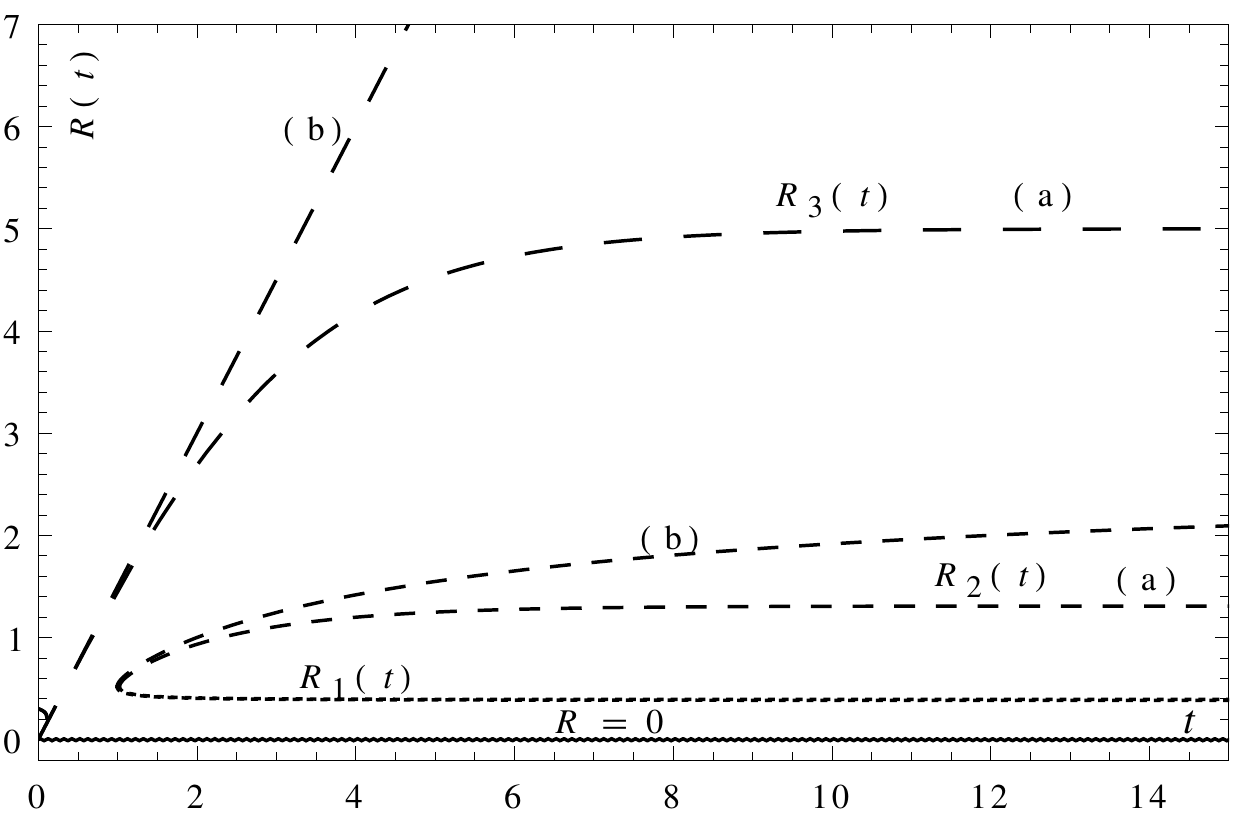}
\caption{The evolution of the apparent horizons as a function of the
time $t$ in the undercharged case of the charged Sultana-Dyer metric
\eqref{chargedSDmetr}.
We used $M=1.5,\, Q=1.0$. The curves (a) are for the scale factor
given by equation~\eqref{factor1} with $k=0.2$, and the curves (b) are for 
$a(t)$
given by equation~\eqref{factor2} with $\alpha =2/3$.
In this case, there is a singularity at $R=0,\,
t=$ finite. For late times, there are three real positive roots of the 
equation
$F_{SD}(R,t) =0$, with $F_{SD}(R,t)$ given by \eqref{F-SD}. }
\label{ap-horizons-Sultana}
\end{center}
\end{figure}

 Looking for singularities in metric~\eqref{chargedSDmetr}  we see that the 
locus
where $g_{tt}=g^{rr} =0$ coincides with the horizons of the
Reissner-Nordstr\"om metric, $r=r_\pm(t)= m(t) \pm \sqrt{m^2(t)-q^2(t)}$,
which are roots
of the function $f(r,t)= 1 -2m(t)/r +q^2(t)/r^2$. However, these are neither
spacetime singularities nor apparent horizons. To find the locus of 
an apparent horizon, we proceed as in the case of
the preceding section and transform to a new radial coordinate given by $R=r
a(t)$. Again, apparent horizons can be
found by analyzing the roots of the equation $g^{RR}(R,t)=0$. In the
case of SD metric \eqref{chargedSDmetr}, this gives
\begin{equation}\label{F-SD}
 F_{SD}(R,t) \equiv H^2(t) R^2 + \left[1 - H(t)\, R\right]^2 g(R,t) -1 =0,
\end{equation}
 where now $g(R,t)=
2M/R - Q^2/R^2$, with $M=m(t)\, a(t)$ and $Q=q(t)\, a(t)$. 
The resulting equation
to be solved is a fourth-order polynomial equation whose zeros depend on
$H(t)$, $M$ and $Q$. For simplicity, we assume here the condition $M^2> 
Q^2$,
with $Q\neq 0$ and consider the two cases for $a(t)$ as earlier, i.e., $a(t)$
given by equations~\eqref{factor1} and \eqref{factor2}.
 For a given range of the parameters $M$
and $Q$, and for the two given choices of $a(t)$, the real roots of
equation~\eqref{F-SD}, $R_i(t)$ ($i=1,\,2,\, ...)$, result in the curves 
drawn in
figure~\ref{ap-horizons-Sultana}.
At early times, there is one real positive root $R_3(t)$ which is a
cosmological apparent horizon (or the Hubble radius) and then the initial
singularity at $R=ra(t)=0$
is not hidden by any kind of horizon. Indeed, the root $R_3(t)$ is
exactly the Hubble time, i.e., the inverse of the Hubble parameter,
$R_3(t) = 1/H(t)$.
At infinitely large times
$R_3(t)$ tends to a constant ($R_3(t)\rightarrow 1/k$) and will form  the de
Sitter cosmological horizon in case (a), i.e., ($R_3,\, t=\infty$) is a de 
Sitter horizon, or it will be the Hubble surface
($R_3(t)=3t/2$) for late times in case (b). For sufficiently large times, two
other solutions $R_1(t)$ and $R_2(t)$ become real and positive. The three
roots exist for all later times and, interestingly, their asymptotic form 
suggest the formation of charged black hole and white hole regions, with a
cosmological horizon. For instance, if $\lim_{t\rightarrow\infty}
H(t)=H_0=$constant, case (a), the three roots tend to fixed values
$R_1(t)\rightarrow
R_{1c}$, $R_2(t)\rightarrow R_{2c}$, and $R_3(t)\rightarrow R_{3c}=1/k$.
We claim that these values correspond respectively to a Cauchy horizon, 
a charged black hole horizon and a de Sitter (cosmological) horizon. The de 
Sitter horizon is defined by ($R_3,\, t=\infty$), the black hole horizon 
$\mathcal{H}$ is given by ($R_2,\, t=\infty$), and the inner horizon is 
defined by $(R_1, \, t=\infty$). 
In the case of matter dominated universe, case (b),
$\lim_{t\rightarrow\infty} H(t)= 0$, $R_1(t)$ and $R_2(t)$ also tend to
constant values with properties of Cauchy and black hole horizons, 
respectively,
while $R_3(t)$ tends to $3t/2$, the Hubble radius. In both cases, the charged
SD metric \eqref{chargedSDmetr} seems to represent a charged black
hole in an expanding universe.

\section{Charged Thakurta type metrics}

\label{sect-thakurta}

\subsection{The solution}

The original Thakurta \cite{thakurta} solution is conformal to the Kerr
metric and was put forward to represent a rotating pointlike mass in an
expanding universe. For the non-rotating case, the Thakurta metric is
conformal to the Schwarzschild metric and is also a candidate to describe the
gravitational field of a point mass in an expanding universe. The
resulting metric in this case was analyzed in \cite{faraoni:2009uy},
even though the
name used there was the SD metric. A charged version of the
non-rotating Thakurta solution was given in \cite{mcclure:2006kg}. 
Such a metric is of the form
\begin{eqnarray}
ds^{2} = -f(r,t) dt^{2}+ a^{2}(t)\left[f^{-1}(r,t)dr^{2}+
r^{2}d\Omega^{2}\right] ,\label{ThKmetric}
\end{eqnarray}
where
\begin{equation}
f(r,t) = 1- \frac{2m(t)}{r} +\frac{q^2(t)}{r^2}\,, \label{f(r)}
\end{equation}
 $m(t)$ and $q(t)$ being related respectively to
the mass and electric charge of the central body, and are fixed parameters 
in 
the original solution.
Metric~\eqref{ThKmetric} may be
obtained from our general metric~\eqref{gen-metric} by choosing
$f_0(r,t)={f_1}^{-1}(r,t)= 1- 2m(t)/r +
q^2(t)/r^2$, $f_2(r,t)=0$, and $f_3(r,t)=1$, so that $g_1(r,t)=1$. As
mentioned above, if $m$ and $q$ are both constant parameters, metric 
\eqref{ThKmetric} is  conformal to the
Reissner-Nordstr\"om metric, as seen by redefining the time coordinate by 
writing $dt = a(\tau)\, d\tau$. 
Despite having been briefly analyzed 
in~\cite{mcclure:2006kg,faraoni:2009uy} (for constant $m$ and constant 
$q$), several aspects of the
resulting metric remain to be investigated. In particular, the sources of the
electromagnetic field deserve further analysis in this case.

 Since we are mostly interested in the
electromagnetic sources, we write here only the metric contributions to the
Einstein tensor which do not depend on $H(t)$,
\begin{equation} \label{ThEinsT} 
{G^t}_t  =  {G^r}_r = - {G^\theta}_\theta = -{G^\varphi}_\varphi =
-\dfrac{q^2(t)}{a^2(t)\,r^4}.
\end{equation}
Note that these are exactly the same as for the charged SD metric
(cf. equations~\eqref{EinstEM-SD}).
These terms should be canceled by the relative components of the 
electromagnetic stress-energy tensor (see next).

\subsection{Faraday and stress-energy tensors, conserved electric charge and
current-density}
\label{sec-EMThK}

Applying the analysis of section~\ref{sec-gen-metric} to the Thakurta
metric~\eqref{ThKmetric} we find the same Faraday-Maxwell tensor as for the
SD case,
$ 
 F^{tr}=-F^{rt} = E(r,t) = {q_0\, h(t)}/({r^2a^3(t)}),$
where, as pointed out above, $q_0$ is a constant and $h(t)$ is an arbitrary
function of the cosmological time $t$ alone, and the only nonzero component
of the current-density is $J^{r} (r,t) = 
-{q_0\,\dot h(t)}/({4\pi\,a^3(t)r^2}) $. 
The electromagnetic stress-energy tensor has the following nonzero terms
\begin{equation}
  E_t^t =  E_r^r =-  E_\theta^\theta=-  E_\varphi^\varphi
=-\dfrac{q_0^2h^2(t)}{8\pi\,a^4(t)r^4 }. \label{Th-emt01}
\end{equation}
Interestingly, the electromagnetic field and the
related stress-energy tensor of the charged Thakurta metric~\eqref{ThKmetric}
are quite the same as for the charged SD
geometry~\eqref{chargedSDmetr}.

The components~\eqref{Th-emt01} should cancel exactly the associated terms in
the Einstein tensor~\eqref{ThEinsT}. Hence, we must choose 
$q_0^2h^2(t)=q^2(t)a^2(t)$. As in the case of the SD metric 
(see  section \ref{sect-SD}), the choice of constant $q(t) =q_0$ as done in
\cite{mcclure:2006kg} implies in $h(t)=a(t)$ and
leads to the presence of a global electric flow in the radial
direction with $J^{r}(r,t)= -q_0H(t)/(4\pi\,a^2(t)r^2)$, and the
total charge of the source varies with time, $Q(t)=q_0a(t)$. 
On the other hand, the choice $Q=\,$constant ($h(t)=1$) has the 
required 
physical property that the total electric charge of the source is constant 
with the cosmological time. This corresponds to 
$q(t)= Q/a(t)$, resulting in a different metric from the solution presented 
in \cite{mcclure:2006kg}.

In conclusion, as in the SD case, there are two interesting cases 
to consider. The first case is the one with $h(t)=a(t)$,  
$q(t)=q_0=$ constant, in which the total charge of the source is $Q(t) = q_0 
a(t)$, and there is a nonzero radial electric current $J^r(r,t)$. As a 
matter of fact, this is the charged Thakurta solution
investigated in \cite{mcclure:2006kg}, but the presence of the nonzero
radial electric current was not reported in that study.
The second case is the one with $h(t) = 1$, $Q(t) = q_0=\,$constant,  and 
$q(t) =  q_0/a(t)$, what leads two a new metric of Thakurta type, which 
is briefly investigated below. 
 
To complete this section let us comment about the changes in the 
energy-momentum tensor implied by the replacement of constant $m$ and $q$ by 
the functions $m(t)=M/a(t)$ and $q(t)=Q/a(t)$. The time-time component of 
the Einstein equations yields ${G^t}_t- 8\pi {E^t}_t 
=-\dfrac{3H^2(t)}{f(r,t)} + \dfrac{2H^2(t)}{f^2(r,t)}\left(\dfrac{M}{a(t)r} 
- \dfrac{Q^2}{a^2(t)r^2}\right) $, where $f(r,t) = 1 - 2M/a(t)r 
+Q^2/a^2(t)r^2$. The first term has the same form as in the 
original Thakurta metric, but now with  $f(r,t)$ replacing $f(r)$ of the 
original solution. 
Considering the source as a fluid with heat flow, we find that the energy 
density of the fluid acquires extra contributions which changes the 
functional form of the energy density. The original energy density is 
positive in all region of the spacetime where $f(r)>0$, while the new 
energy density becomes negative for values of $R=a(t)r$ very close to the 
largest zero of the function $f(r,t)$. Similar changes occur in the 
pressure and heat flow, but we do not show such functions here. This 
negative energy density may be interpreted as a kind of phantom matter. On 
the other hand, we can avoid such a situation by 
considering, for instance, a mixture of fluids instead of a single fluid 
with heat flow. Other interesting way out of this problem is to take 
different functions for 
$m(t)$ while keeping $q(t) = Q/a(t)$, which is the appropriate 
electromagnetic source. To simplify the analysis and for the sake of 
comparison 
with the McVittie metric, we treat here the case $m(t)=M/a(t)$ only.

\subsection{Singularities and horizons of the charged Thakurta
metric}

 As in the previous cases, for
simplicity, we comment only on the undercharged case.

For constant $m$ and constant $q$, the Ricci 
scalar for the metric~\eqref{ThKmetric} may be written as  $
 {\cal R }= {12H^2(t)}/{f(r)}+ {6 \dot{H}(t)}/{f(r)},$
and the Kretschmann scalar is given in \ref{appendixTh}.
The other curvature scalars present the same singularities as those found
from Ricci \eqref{ThRicci} and Kretschmann \eqref{ThKretsch}.

As seen from the Ricci scalar there are singularities at regions of the
spacetime where at least one of the following conditions is satisfied:
when $H(t)$ diverges, when $a(t)$ vanishes, and where $f(r)=0$.
Thence, besides a possible big-bang singularity when $a(t)\rightarrow 0$
($H(t)\rightarrow \infty$), the singularities are defined by
the zeros of function $f(r)= 1 -2m/r + q^2/r^2$, i.e., at $r=r_\pm = m \pm
\sqrt{m^2 +q^2}$. Of course, the existence of such singularities depends upon
the asymptotic behavior of the Hubble parameter $H(t)$ for
$t\rightarrow\infty$. For instance, if $\lim_{t\rightarrow\infty}
H(t) =0$ and $\lim_{t\rightarrow\infty} \dot H(t) =0$, then the singularity
at $f(r)=0$ that exists for finite $t$ disappears in the infinite future of
time $t$.

Furthermore, from the Kretschmann scalar (and from other curvature scalars
such as $R_{\mu\nu} R^{\mu\nu}$) it follows that $r=0$ is also a singularity.
Hence, the singularity is at $r=0$ in the case $f(r)$ has no real roots,
i.e., in the overcharged case, where the central object has electric charge
$q$ larger than the mass parameter $m$, $m^2< q^2$. 

The same singularities are present for non-constant $m(t)$ and $q(t)$. 
The only difference is that the singularities defined by the zeros 
of the function $f(r,t)=1- 
2m(t)/r +q^2(t)/r^2$ now depend on time. The curvature scalars for this 
case are not written here since their expressions
are cumbersome (see \ref{appendix}).

 As done in the previously analyzed spacetimes,
it is useful to define a physical radial coordinate by $R=r\,a(t)$,
so that metric \eqref{ThKmetric} assumes the form
\begin{equation}\label{Thmetric2}
\begin{split}
 ds^{2}=&-\left(f(R)-\frac{H^{2}(t)\,R^2}{f(R)}\right)dt^{2}
-\frac{2H(t)\, R}{f(R)}\,dtdR+\frac{dR^{2}}{f(R)}  +R^{2}d\Omega^{2}, 
                                 \end{split}
\end{equation} 
where $f(R)= 1-{2M}/{R} + {Q^2}/{R^2}$ and the new parameters
$M=m(t)\,a(t)$ and $Q=q(t)\,a(t)$ were defined.
Again, the search for apparent horizons is performed just in the case 
$m(t) = M/a(t)$ and $q(t) = Q/a(t)$, with constant $M$ and $Q$.

In terms of the new parameters $M$ and $Q$, for $M^2\geq Q^2$ the curvature
singularities are located at loci of the spacetime where $f(R)=
1-{2M}/{R} + {Q^2}/{R^2} =0$, i.e.,
$ R_\pm = M\pm
\sqrt{M^2 - Q^2},$
where $R_\pm = a(t)\, r_\pm = a(t)\left(m\pm \sqrt{m^2 -q^2}\right)$.
In the overcharged case, $M^2<Q^2$, there is a curvature singularity at
$R=0$.

Since there are singularities in the generalized Thakurta spacetime given by 
metric~\eqref{Thmetric2}, it is interesting
to verify whether there are horizons hiding the singularity to
external observers, so we look for apparent horizons. In terms of the
metric~\eqref{Thmetric2}, apparent horizons, if exist, are given by the 
roots of the equation $H^2(t)\, R^2 - f^2(R)=0$, or
\begin{equation}\label{F-Th}
F_{T}(R,t) = H^2(t)\,R^2 -\left(1-\frac{2M}{R}+\frac{Q^{2}}{R^{2}}\right)^2
= 0\, , 
\end{equation}
which results in a third-order polynomial equation for $R$.

The analytical expressions for the three roots of the above polynomial
equation \eqref{F-Th} for $R$ ($R_1(t)$, $R_2(t)$, and $R_3(t)$) are not
written explicitly here. The solutions depend explicitly on the Hubble
parameter $H(t)$. Restricting the
analysis to expanding universes $H(t)\geq 0$, so that $H(t)\, R > 0$, the
nature of such roots depends on the charge to mass relation. 
A deeper analysis shows that for $M^2\geq Q^2$ one of the roots, $R_1(t)$,
say, is smaller than $R_{-}=M-\sqrt{M^{2}-Q^{2}}$ for all times, while the
other two are larger than $R_{+}=M+\sqrt{M^{2}-Q^{2}}$. On the other hand,
for $M^2< Q^2$ there is only one real positive root of 
equation~\eqref{F-Th}, 
but as mentioned above
we report here mostly on the undercharged case ($M^2 > Q^2$). The apparent
horizons of the extremely charged Thakurta metric ($M^2=Q^2$) have a similar
time behavior as for the undercharged case.

\begin{figure}[!ht]
\begin{center}
\includegraphics[width=3.1in]{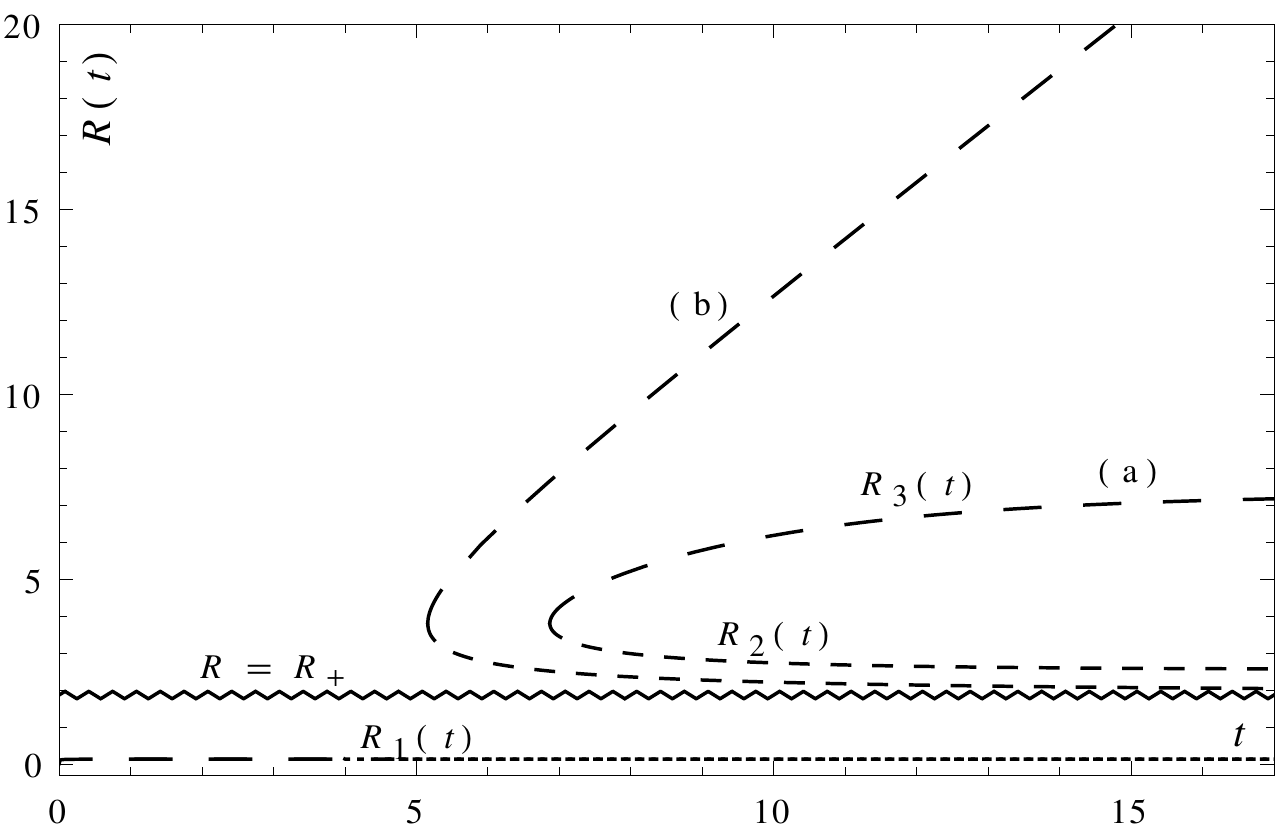}
\caption{The evolution of the apparent horizons as a function of the
time $t$ in the undercharged case of the charged Thakurta
spacetime~\eqref{Thmetric2},  for two different expansion factors, as
in the Vaidya (figure~\ref{horizons-Vaidya}) and SD
(figure~\ref{ap-horizons-Sultana}) metrics. The plots are for $M=1.0$, 
$Q=0.5$,
$a(t)$ given by equation~\eqref{factor1} with $k=0.1$ for case (a), and 
$a(t)$
given by equation~\eqref{factor2} with $\alpha = 2/3$ for case (b).
The initial singularity at $R=R_+$, $t$ finite, is indicated. For late 
times, there are three real positive roots of equation
$F_{T}(R,t) =0$, with $F_T(R,t)$ given by \eqref{F-Th}.}
\label{horizons-Thakurta-under}
\end{center}
\end{figure}

 If $M^2\geq Q^2$ there is a curvature
singularity at $R=R_+$, with finite  $t$, and
the three roots may be real and positive. According to the Big-Bang
cosmological models, $H(t)$ is large at early times and then, for small 
times, equation~\eqref{F-Th} has only one real positive root ($R_1(t)$), the 
other two
roots being complex conjugate to each other. As time goes on, $H(t)$
decreases and there is a specific time $t_b$ for which $
{27\left(-1+6MH(t_b)-Q^{2}H^{2}(t_b)\right)} =
{2\left(1-6MH(t_b)\right)^{3/2}}$ when these roots become real and
equal to each other, $R_2(t_b)=R_3(t_b)$. Thereafter, for all later
times, the three horizons develop independently. The situation is depicted in
figure~\ref{horizons-Thakurta-under}, where we plot $R_+$ and the three
solutions of
equation~\eqref{F-Th} as a function of the time coordinate $t$ for two cases 
of
$a(t)$.  At early times, the
singularity $R=R_+$ ($t$ finite) is completely naked, and after some
intermediate time $t=t_b$ an apparent horizon is formed at
$R=R_2(t_b)=R_3(t_b)$; it immediately bifurcates into two branches that
evolves independently.
 
 The solution $R_2(t)=R_3(t)$, which is satisfied at a particular time,
bifurcates and after that $R_2(t)$ and $R_3(t)$ evolve forming 
a cosmological apparent horizon (or Hubble type surface) $R_3(t)$, and
a black hole type apparent horizon $R_2(t)$. Asymptotically (for
$t\rightarrow\infty$), in the
case (a) where $\lim_{t\rightarrow\infty} H(t)=H_0>0$, $R_3(t)$ tends
to a de Sitter type cosmological horizon, 
$\lim_{t\rightarrow\infty} R_3(t)=R_{3c}=$
constant in time ($R_{3c}$ is the largest real root of the equation
$1-2M/R+Q^2/R^2 - H_0 R =0$). In the case (b), where  
$\lim_{t\rightarrow\infty} H(t)= 0$,  $R_3 (t)$ tends to the 
Hubble radius, $R_3(t)\sim 1/t$. 
On the other hand, in both cases (a) and (b), $R_2(t)$  tends
to a black hole horizon. In the case (a) $\lim_{t\rightarrow\infty} R_2(t)=
R_{2c}=$ constant (which is the intermediate solution of the equation
$1-2M/R+Q^2/R^2 - H_0 R =0)$, while in case (b)  $R_2(t)$ tends to the
Reissner-Nordstr\"om black hole horizon $R_+= M + \sqrt{M^2 - Q^2}$.   
Following \cite{lake}, see also \cite{fontanini}, we see that the 
region ($R_2,\ t=\infty$) is a black hole horizon $\mathcal{H}$.
Notice also that, in the limit $t\rightarrow\infty$, for case (b) the Ricci 
scalar tends to zero and
the regions of the spacetime
$R_+$ and $R_-$ are no longer curvature singularities. For this particular
case, the true singularity is at $R=0$, as it follows from the Kretschmann 
scalar~\eqref{ThKretsch}. With this we may infer that the boundaries of the
resulting spacetime are similar to the Reissner-Nordstr\"om case.

Lets stress also that if the universe has a phase
dominated by the cosmological constant (at late times) the Hubble parameter
$H(t)$ tends to a constant, and  $R_2(t)$ tends to a constant value larger
than $R_+$, i.e., $\lim_{t\rightarrow\infty} R_2(t)  = R_{2c}=$
constant, and thus a black hole horizon is formed. 
The cosmological-like horizon $R_3(t)$ tends to a de Sitter 
horizon (see  \cite{lake}). In this case we may infer that the causal 
structure of the region $R>R_+$ of the charged Thakurta type spacetime 
\eqref{Thmetric2}
is similar to the McVittie metric~\cite{lake,fontanini}.

\section{Further comments and conclusion}
\label{sect-conclusion}

The McVittie metric was proposed in 1933 and have being studied by many
authors. Though, we can say that a good comprehension of the
global structure of the corresponding spacetime has been achieved only
recently. It is then clear that a lot of
effort is still needed to reach a good level of understanding on the 
other metrics that have been written to represent black
holes in cosmological backgrounds. This is so for the Vaidya metric
\cite{vaidya}, and also for the Sultana-Dyer \cite{sultana} and Thakurta
\cite{thakurta} metrics. What we mean is that not even the original
versions of such metrics were investigated in detail, and so any progress in
the charged cases is of high interest.

 We started our study reviewing the charged metrics given
in \cite{mcclure:2006kg}, and soon verified that the 
electromagnetic field of those solutions deserved further examination,
since in some cases the correct sources of the
Maxwell electromagnetic field were not given. In particular, in the charged
versions of the SD and Thakurta metrics the presence of a radial
current density across the spacetime was not observed. Then, by the sake of
comparison, we started by investigating this issue in the charged McVittie
solution given in \cite{vaidya2}, and also considered a charged
Vaidya type metric. In both cases, Maxwell
equations were solved by choosing the Faraday-Maxwell field in such a
way that the four-current density vanishes everywhere except at singular
points (regions) of the spacetime. The total charge of the central object is 
a constant, and the stress-energy tensor emerging from such a
Faraday-Maxwell gauge field exactly cancels the corresponding terms in the
Einstein tensor produced by the given metric. On the other hand, in the
charged versions of the SD and Thakurta type metrics, the 
situation is a
little more interesting. The Faraday-Maxwell tensor field related to zero
four-current outside the source contributes to a stress-energy tensor which
does not cancel the corresponding terms of the Einstein tensor coming from
the metrics given in \cite{mcclure:2006kg}. This gives rise to
additional energy and pressure terms that have no direct physical
interpretation. On the other hand, if one chooses the correct
Faraday-Maxwell field so to get rid of the undesirable energy terms, then
there is a global (radial) electric current throughout the spacetime, and the
electric charge of the source varies with the
cosmological time. 

 As commented above, further studies are necessary to complete the analysis
of the metrics considered here. As we have shown, some of them present
properties that may represent charged black holes in expanding universes.
It is worth mentioning once more that the geodesic structure of most of those
spacetimes were not investigated in the literature, not even of the 
uncharged original metrics. We are now investigating the global structure of 
the charged SD and Thakurta metrics. The complete analysis, 
including the
conformal diagrams, will be published in separate papers.

\section*{Acknowledgments}

M.G.R. thanks Coordena\c{c}\~ao de Aperfei\c{c}oamento de Pessoal de N\'\i
vel Superior - CAPES, Brazil, for a grant (Processo 2010/059582).
V.T.Z. would like to thank
Conselho Nacional de Desenvolvimento Cient\'ifico e Tecnol\'ogico -
CNPq, Brazil, for grants, and Funda\c{c}\~ao de Amparo \`a Pesquisa do
Estado de S\~ao Paulo for a grant (Processo 2012/08041-5).

\appendix

\section{Curvature scalars}
\label{appendix}

\noindent
{\bf a) Charged McVittie metric}:
\label{appendixMcV}

The charged McVittie metric gets a simpler form by using a
Schwarszchild-like radial coordinate $R$ given by
\begin{equation}
\hspace*{-.2cm}   R = a(t)\,r\,g(r,t) = a(t)\, r\left(1 +\frac{m}{2a(t)\, r}
\right)^2
- \dfrac{q^2}{4a(t)\, r}.
\end{equation}
In terms of this new radial coordinate, and taking constant mass and 
charge parameters, $m=M$ and $q=Q$, the Ricci and the Kretschmann
scalars, ${\cal R}$ and ${\cal K}$, respectively, for the charged McVittie
metric~\eqref{chargedMcVmetr} may be written as
\begin{eqnarray}
 &&\hspace*{-2.3cm} {\cal R}=12 H^2(t)+ \dfrac{6\dot H(t)}{\sqrt{f(R)}},\\
 &&\hspace*{-2.3cm} {\cal K}=  48
\left(\dfrac{M}{R^3}-\dfrac{Q^2}{R^4}\right)^2 +
\frac{8Q^4}{R^8} + 24H^4(t) 
   + \frac{4\dot{H}(t)}{\sqrt{f(R)}}
  \left( 3\dot{H}(t) + {6H^2(t)}- \dfrac{2Q^2}{R^4}\right),
\label{McVKretsch}
\end{eqnarray}
where we defined $f(R) = 1-2M/R + Q^2/R^2.$
These curvature scalars are used in the study performed in
section~\ref{sect-mcvittie}.\\

\noindent{\bf b) Charged Vaidya type metrics}
\label{appendixVy}

The Ricci and the Kretschmann scalars, ${\cal R}$ and ${\cal K}$,
respectively, for the Vaidya metric~\eqref{chargedvaidyametr}, considering 
$m=\,$constant and $q=\,$constant,  are
given by 
\begin{eqnarray}
\hspace*{-2.3cm}
{ \cal R} &=&2H^{2}(t)\left(6+\frac{g(r)}{a^2(t)}\right)+2\dot{H}(t)\left(3+
\frac{2g(r)}{a^2(t)}\right)-\frac { 4mH(t)}{a^{3}(t)r^{2}}, \label{ricciVd}\\
\nonumber
\hspace*{-2.3cm}
{\cal K} &=&12\left(2H^{4}(t)+2H^{2}(t)\dot{H}(t)+\dot{H}^{2}(t)\right) +
\frac{8g(r)}{a^2(t)}\left(H^{4}(t)+2H^{2}(t)\dot{H}(t)+2\dot{H}^{2}(t)\right)
 \label{kretschVd}
 \\ \nonumber
 \hspace*{-2.3cm}
& + &\frac{g^2(r)}{a^2(t)}\left(8\dot{H}^2(t) +H^{4}(t)
-\frac{16H^{3}(t)}{a(t)r}\right)
 - \frac{16{g(r)}\dot{H}(t)}{a^{6}(t)r^{4}}\left(q^{2}
-mr\right)\left[1-a(t)rH(t)\right]
\\ \nonumber
\hspace*{-2.3cm}
& -& \frac{8\dot{H}(t) q^{2}}{a^{4}(t)r^{4}}+\frac{16mH(t)}
{a^{3}(t)r^{2}}\left(\dot{H}(t)-H^{2}(t)\right)
+\frac{8H^{2}(t)}{a^{6}(t)r^{2}}\left(\frac{7q^{4}}{r^4}
-\frac{20mq^{2}}{r^3}+\frac{16m^2}{r^2} \right) \\  
\hspace*{-2.3cm}
&-&\frac{16H(t)}{a^{7}(t)r^{3}}\left(\frac{6q^{4}}{r^4}-\frac{13mq^{2}}{r^3}
+\frac{8m^2}{r^2} \right)
+\frac{8}{a^{8}(t)r^{4}}\left(\frac{7q^{4}}{r^4}-\frac{12mq^{2}}{r^3}+
\frac{6m^{2}}{r^2} \right), 
\end{eqnarray} 
where $g(r)=\dfrac{2m}{r} -\dfrac{q^2}{r^2}$.   These scalars are used in the
analysis of the
singularities of the charged Vaidya spacetime, in section~\ref{sect-vaidya}.

Now, taking time dependent mass and charge parameters, i.e., with $m=m(t)$ 
and $q=q(t)$,   besides the 
terms showed by equation~\eqref{ricciVd} and \eqref{kretschVd} with $m$ and 
$q$ 
replaced by $m(t)$ and $q(t)$, the resulting Ricci and Kretschmann scalar 
acquire a large number of additional terms. Thus, 
to avoid cumbersome expressions in the paper, we do not write them here. 
Such extra terms contain negative powers of $a(t)$, positive powers $H(t)$ 
and $\dot{H}(t)$, and negative powers of $r$. These extra terms contain also 
positive powers of $m(t)$, $q(t)$, $\dot{m}(t)$, and $\dot{q}(t)$.

If we take $m(t) = M/a(t)$ and $q(t)=Q/a(t)$, with constant $M$ and $Q$, as 
considered in the text, the Ricci scalar becomes 
\begin{eqnarray}
{ \cal R} &=&12H^{2}(t)+3\dot{H}(t)\left(2+
\frac{g(r,t)}{a^2(t)}\right)+\frac {\dot{H}(t)Q^2}{a^{4}(t)r^{2}}, 
\label{ricciVd2} \hskip .5cm
\end{eqnarray}
where $g(r,t)=\dfrac{2M}{a(t)r} -\dfrac{Q^2}{a^2(t)r^2}$.
The Kretschmann scalar results a too long expression to be written here, but 
it has the same kind of terms as in the case of constant $m$ and $q$.  
Hence, the singularities are 
the same as in the case with constant $m$ and $q$. Namely, a big-bang type
singularity when $a(t) =0$, or, equivalently, when $H(t)$ and $\dot{H}(t)$ 
become infinitely large, and a singularity at the region ($r=0,\, t$ finite).
\\

\noindent{\bf c) Charged Sultana-Dyer type metrics}
\label{appendixSD}

For constant $m$ and $q$, the Ricci and the Kretschmann scalar obtained from
metric~\eqref{chargedSDmetr} are, respectively,
\begin{eqnarray} \label{RicciSD}
{\cal R} =  6\left(2H^{2}(t)+ \dot{H}(t)
\right)\left[1+ g(r)\right]
-\frac{12mH(t)} {a(t)r^{2}},
\end{eqnarray}  
\begin{eqnarray}
\hspace*{-2.3cm}{\cal K}
&=& 12\left(1+g(r)\right)^{2}\left(2H^{4}(t)+2H^{2}(t)\dot{H}(t)+\dot{H}^{2}
(t)\right) + \frac{16q^{2}H(t)}{a^{3}(t)r^{7}}\left(2q^{2}-3mr\right)
\nonumber\\
\hspace*{-2.3cm}&&
-\frac{16\dot{H}(t)
H(t)}{a(t)r^{3}}\left(q^{2}g(r)+mr\right)
-\frac{8\dot{H}(t)}{a^{2}(t)r^{4}}\left(1+g(r)\right)
-\frac{48mH^{3}(t)}{a(t)r^{2}}\left(1+g(r)\right)\nonumber  \\
\hspace*{-2.3cm}& &
+\frac{32H^{2}(t)}{a^{2}(t)r^{6}}\left(q^{4}-3mq^{2}r+3m^{2}r^{2}\right)+
\frac{8}{a^{4}(t)r^{8}}\left(7q^{4}-12mq^{2}r+6m^{2}r^{2}\right), 
\end{eqnarray} \label{KretSD}
where $g(r)=\dfrac{2m}{r} -\dfrac{q^2}{r^2}$. 

The modifications of these scalars in the case $m=M/a(t)$ and $q=Q/a(t)$, 
with constant $M$ and $Q$, correspond to the presence of new 
terms with the same functional dependence on $a(t)$, $H(t)$ , $\dot{H}(t)$, 
and $r$ as the above terms. For instance, the Ricci scalar changes to 
$ 
{\cal R} =  6H^{2}(t)\left(2 + \dfrac{2M}{a(t)r} - 
\dfrac{Q^2}{a^2(t)r^2}\right)+ 2\dot{H}(t)\left(3 +\dfrac{5M}{a(t)r} 
-\dfrac{Q^2}{a^2(t)r^2}\right)
-\dfrac{8MH(t)} {a(t)r^{2}}$.
We do not write the resulting expression for the Kretschmann scalar
because it is too cumbersome.
These scalars 
are used in the analysis of singularities of the charged SD
spacetime in section~\ref{sect-SD}. \\

\noindent{\bf d) Charged Thakurta type metrics}
\label{appendixTh}

Taking constant parameters $m$ and $q$, the Ricci and the Kretschmann 
scalars of the charged Thakurta 
metric~\eqref{ThKmetric} may be 
written respectively as 
\begin{equation} \label{ThRicci}
 {\cal R }= \dfrac{12H^2(t)}{f(r)}+ \dfrac{6 \dot{H}(t)}{f(r)},
\end{equation}
\begin{eqnarray}\label{ThKretsch}
 {\cal K} &= &
\frac{12}{f^2(r)}\left(\dot H(t) + H^2(t)\right)^2 -\frac{8q^2\, \dot
H(t)}{a^2(t)\,r^4\,f(r)} + 
\frac{12\, H^4(t)}{f^2(r)}\nonumber \\ 
&&\hspace{-.6cm} +
\left(\dfrac{m}{r^3}-\dfrac{q^2}{r^4}\right)^2
 \left(\dfrac{48}{a^4(t)}-\frac{16H^2(t)}{a^2(t)\,f^2(r)} \right)+
\dfrac{8q^4}{a^4(t)\, r^8} .
\end{eqnarray}
These scalars are used in the analysis of section~\ref{sect-thakurta}.

As in the SD metric, replacing $m$ and $q$ by function 
of time $m(t)$ and $q$ does note affect the big-bang singularity $a(t)=0$, 
or $H(t)=\infty$ and $\dot{H}(t)=\infty$. The singularities at the zeros of 
the function $f(r,t)=1-2m(t)/r+q^2(t)/r^2$ are also present, but now depend 
on 
the cosmic time $t$. We do not write here the expressions for the Ricci and 
Kretschmann scalars for the general because they are too large.  
 In the 
particular case where $m(t)= M/a(t)$ and $q(t)=Q/a(t)$, the Ricci scalar 
becomes  ${\cal R }= {12H^2(t)}/{f(r,t)}- {2\dot{H}(t)}
\left( 6f(r,t) +{M}/{a(t)r} -{Q^2}/{a^2(t)r^2}\right)/{f^2(r,t)}
+ {2H^2(t)}\left[4\left({M}/{a(t)r} - 
{Q^2}/{a^2(t)r^2} \right)^2 - 
f(r,t)\left({6M}/{a(t)r}-{5Q^2}/{a^2(t)r^2}\right)\right]/ 
{f^3(r,t)}
$, and 
we avoid to write the huge expression for the Kretschmann scalar.

 \end{document}